\documentclass[10pt]{article}
\usepackage{a4wide}
\usepackage{cite}
\usepackage{amsmath}
\usepackage{amsfonts}
\usepackage{graphicx}

\newcommand{\dd}{{\rm{d}}}        
\newcommand{\sign}{{\rm sign}}    
\newcommand{\A}{{\mathcal{A}}}    
\newcommand{\T}{{\mathcal{T}}}    
\newcommand{\HH}{{\mathcal{H}}}   
\newcommand{\PP}{{\mathcal{P}}}   
\newcommand{\QQ}{{\mathcal{Q}}}   
\newcommand{\scri}{{\mathcal{I}}} 

\newcommand{\sdfrac}[2]{\mbox{\small$\displaystyle\frac{#1}{#2}$}}

\def \bF {\mathbf{F}}
\def \bA {\mathbf{A}}

\newcommand{\rovno}{\!\!\!\!& = &\!\!\!\!}
\newcommand{\equi}{\!\!\!\!& \equiv &\!\!\!\!}

\begin{document}

\title{New form of all black holes of type D\\ with a cosmological constant}

\author{
Ji\v{r}\'{\i} Podolsk\'y
and
Adam Vr\'atn\'y\thanks{
{\tt podolsky@mbox.troja.mff.cuni.cz}
and
{\tt vratny.adam@seznam.cz}
}
\\ \ \\ \ \\
Charles University, Faculty of Mathematics and Physics, \\
Institute of Theoretical Physics, \\
V~Hole\v{s}ovi\v{c}k\'ach 2, 18000 Prague 8, Czechia.
}

\maketitle

\begin{abstract}
We present an improved metric form of the complete family of exact black hole spacetimes of algebraic type~D, including any cosmological constant. This class was found by Debever in 1971, Pleba\'nski and Demia\'nski in 1976, and conveniently reformulated by Griffiths and Podolsk\'y in 2005. In our new form of this metric the key functions are simplified, partially factorized, and fully explicit. They depend on seven parameters with direct physical meanings, namely $m, a, l, \alpha, e, g, \Lambda$ which characterize mass, Kerr-like rotation, NUT parameter, acceleration, electric and magnetic charges of the black hole, and the cosmological constant, respectively. Moreover, this general metric reduces directly to the familiar forms of  (possibly accelerating) Kerr--Newman--(anti-)de Sitter spacetime, charged Taub--NUT--(anti-)de Sitter solution, or (possibly rotating and charged) $C$-metric with a cosmological constant by simply setting the corresponding parameters to zero. In addition, it shows that the Pleba\'nski--Demia\'nski family does not involve accelerating NUT black holes without the Kerr-like rotation. The new improved metric also enables us to study various physical and geometrical properties, namely the character of singularities, two black-hole and two cosmo-acceleration horizons (in a generic situation), the related ergoregions, global structure including the Penrose conformal diagrams, parameters of cosmic strings causing the acceleration of the black holes, their rotation, pathological regions with closed timelike curves, or thermodynamic quantities.
\end{abstract}

\vfil\noindent
PACS class:  04.20.Jb, 04.70.Bw, 04.40.Nr, 04.70.Dy



\vfil
\eject

\section{Introduction}
\label{intro}

Black holes belong to the most remarkable predictions of Einstein's general relativity. Although their existence had been doubted for many decades, it is now widely accepted that such \emph{totaly} gravitationally collapsed ``objects'' indeed exist in our Universe. Recent (and spectacular) observational proofs of this fact are the detections of gravitational waves emitted from binary black hole coalescences, achieved by the LIGO Scientific Collaboration--Virgo Collaboration \cite{LIGO-Virgo:2016, LIGO-Virgo:2021}, and also the first direct image of a shadow of a supermassive black hole in M87* and in Sgr~A*, obtained by the Event Horizon Telescope Collaboration \cite{EHT:2019, EHT:2022}.

First \emph{exact spacetimes} representing black holes were found very soon after the final formulation of Einstein's field equations of general relativity in November 1915. Namely, it is the important solution of Schwarzschild (1916), Reissner--Nordstr\"{o}m solution with an electric charge (1916--1918), and Kottler--Weyl--Trefftz solution with a cosmological constant $\Lambda$ (1918--1922). These were followed in 1960s by rotating Kerr (1963), twisting Taub--NUT (1963) or Kerr--Newman charged black holes (1965), and also the so called $C$-metric (1918, 1962), physically interpreted by Kinnersley--Walker (1970) as uniformly accelerating pair of black holes.

All these fundamental exact solutions are spherically/axially symmetric, and are of algebraic type~D. In fact, they belong to a general family of type~D spacetimes with any~${\Lambda}$ and an aligned electromagnetic field. Nonaccelerating solutions of this family were obtained in 1968 by Carter \cite{Carter:1968b}. In the vacuum ${\Lambda=0}$ case, they include all the particular subclasses identified by Kinnersley \cite{Kinnersley:1969}. Debever \cite{Debever:1971} in 1971 found a wider class of such black holes which also admit acceleration. In 1976 a better metric representation of this complete class of type~D exact solutions to Einstein--Maxwell equations with double-aligned non-null electromagnetic field and~${\Lambda}$ was found in a seminal work \cite{PlebanskiDemianski:1976} by Pleba\'nski and Demia\'nski (for more details and further references see \cite{Stephanietal:2003} and \cite{GriffithsPodolsky:2009}, in particular Chapter~16).

Unfortunately, the familiar forms of the well-known black holes \emph{were not included explicitly} in the original Pleba\'nski--Demia\'nski metric (specific degenerate transformations had to be applied), and the \emph{physical interpretation} of its seven free parameters was not clear. Both these drawbacks were overcome in 2006 in the works of Griffiths and Podolsk\'y \cite{GriffithsPodolsky:2005,GriffithsPodolsky:2006,PodolskyGriffiths:2006}, see also \cite{GriffithsPodolsky:2009}, enabling easier analysis of physical and geometrical properties of these exact black holes.

In our recent paper \cite{PodolskyVratny:2021} we demonstrated that this Griffiths--Podolsk\'y form of the generic black-hole metric of type~D can be \emph{further improved}. This was achieved by introducing a modified set of the mass and charge parameters, an appropriate conformal rescaling, and a useful gauge choice of the twist parameter. The new improved form of the metric is simple, fully explicit, and with factorized metric functions. It is thus possible to investigate and evaluate  various properties of this large family of rotating, charged, and accelerating black holes, namely their singularities, horizons, ergoregions, infinities, cosmic strings, or thermodynamics \cite{PodolskyVratny:2021}.

In such studies we restricted ourselves only to the case ${\Lambda=0}$. It is the purpose of the present paper to extend the new improved coordinate representation found in \cite{PodolskyVratny:2021} to \emph{any value of the cosmological constant}, thus completing our program to improve the metric description of the \emph{full class} of Pleba\'nski--Demia\'nski black holes of algebraic type~D.

In Sec.~\ref{sec_derivation} we systematically derive the new form of the metric, with the results summarized in Sec.~\ref{sec:introsummary}. In subsequent Sec.~\ref{sec_subclasses} all the main subclasses of this large family of type~D black holes are discussed --- these are obtained by simply setting the corresponding physical parameters $\Lambda, \alpha, l, a, e, g$ to zero. The second part of our paper, which is contained in the long Sec.~\ref{sec_discussion}, is devoted to the physical and geometrical analysis of this class of black holes which can be done fully explicitly using our improved form of the generic metric. Such a study includes determining the curvature of the gravitational field, evaluation of the electromagnetic field, the structure and location of horizons, finding the related ergoregions, analytic extension and global structure, regularization of the symmetry axes, properties of the possible cosmic strings or struts, their rotation related to the NUT parameter, regions with closed timelike curves in their vicinity, and calculation of the entropy and temperature of the black-hole and cosmo-acceleration horizons. Final summary with further remarks is contained in Sec.~\ref{sec:summary}.

\section{Derivation of the new form of the metric}
\label{sec_derivation}

First, let us recall the convenient representation of the complete class of Pleba\'nski--Demia\'nski black holes of algebraic type~D found by Griffiths and Podolsk\'y in 2005 \cite{GriffithsPodolsky:2005,GriffithsPodolsky:2006,PodolskyGriffiths:2006}. It is summarized in Eq.~(16.18) of \cite{GriffithsPodolsky:2009} as
\begin{align}
\dd \tilde{s}^2 = \frac{1}{\Omega^2} &
  \bigg(\!-\frac{\QQ}{\rho^2}\big[\,\dd t- \big(a\sin^2\theta +4l\sin^2\!\tfrac{1}{2}\theta \big)\dd\varphi \big]^2
   + \frac{\rho^2}{\QQ}\,\dd r^2  \nonumber\\
& \quad  + \,\frac{\rho^2}{\PP}\,\dd\theta^2
  + \frac{\PP}{\rho^2}\,\sin^2\theta\, \big[ a\,\dd t -\big(r^2+(a+l)^2\big)\,\dd\varphi \big]^2
 \bigg), \label{newmetricGP2005-0}
\end{align}
where the metric functions are
\begin{eqnarray}
\Omega    \rovno 1 - \frac{\alpha}{\omega}\, (l+a \cos\theta )\, r\,, \label{Omega}\\[1mm]
\rho^2    \rovno r^2+(l+a \cos \theta)^2 \,, \label{rho}\\[2mm]
\PP(\theta) \rovno 1-a_3\cos\theta-a_4\cos^2\theta\,,  \label{P}\\
\QQ(r)    \rovno (\omega^2k+\tilde{e}^2+\tilde{g}^2)-2\tilde{m}\,r+\epsilon\,r^2-2\alpha\,\frac{n}{\omega}\,r^3
                  -\Big(\alpha^2k+\frac{\tilde{\Lambda}}{3}\Big)\,r^4. \label{Q}
\end{eqnarray}
The constants $a_3$ and $a_4$ in \eqref{P} are
\begin{align}
a_3 &= 2\alpha \,\frac{a}{\omega}\,\tilde{m} -4\alpha^2\, \frac{a\,l}{\omega^2}\,  (\omega^2k+\tilde{e}^2+\tilde{g}^2)
    -4\,\frac{\tilde{\Lambda}}{3}\,a\,l\,,  \label{a3}\\
a_4 &= -\alpha^2 \frac{a^2}{\omega^2}\,(\omega^2k+\tilde{e}^2+\tilde{g}^2) -\frac{\tilde{\Lambda}}{3}\,a^2\,, \label{a4}
\end{align}
while the coefficients $\epsilon$, $n$, and $k$ in \eqref{Q}--\eqref{a4} are determined by the relations
\begin{align}
\epsilon &= \frac{\omega^2k}{a^2-l^2} +4\alpha\,\frac{l}{\omega}\,\tilde{m}
 -(a^2+3l^2) \Big[\, \frac{\alpha^2}{\omega^2}\,(\omega^2k+\tilde{e}^2+\tilde{g}^2)+\frac{\tilde{\Lambda}}{3}\, \Big],
  \label{epsilon}\\
n &= \frac{\omega^2k}{a^2-l^2}\,l -\alpha\,\frac{a^2-l^2}{\omega}\,\tilde{m}
 +(a^2-l^2)\,l\, \Big[\, \frac{\alpha^2}{\omega^2}\,(\omega^2k+\tilde{e}^2+\tilde{g}^2)+\frac{\tilde{\Lambda}}{3} \Big],
  \label{n}
\end{align}
and
\begin{align}
 \Big(\, \frac{\omega^2}{a^2-l^2} + 3\alpha^2\,l^2 \Big) \,k
    = 1 +2\alpha\,\frac{l}{\omega}\,\tilde{m} -3\alpha^2\frac{l^2}{\omega^2}(\tilde{e}^2+\tilde{g}^2)
     -\tilde{\Lambda}\, l^2,
  \label{k}
\end{align}
which implies
\begin{align}
  \frac{\omega^2 k}{a^2-l^2} &=
  \frac{{\displaystyle 1-\tilde{\Lambda}\, l^2 +2\alpha \frac{l}{\omega}\,\tilde{m} -3\alpha^2\frac{l^2}{\omega^2}\,(\tilde{e}^2+\tilde{g}^2)}}
  {{\displaystyle 1 + 3\alpha^2 \frac{l^2}{\omega^2}\,(a^2-l^2)}}\,,
  \label{kfrac} \\
  (\omega^2 k+\tilde{e}^2+\tilde{g}^2) &=
  \frac{{\displaystyle (1 -\tilde{\Lambda}\, l^2)(a^2-l^2) + (\tilde{e}^2+\tilde{g}^2) +2\alpha \frac{l}{\omega}(a^2-l^2)\,\tilde{m} }}
  {{\displaystyle 1 + 3\alpha^2\frac{l^2}{\omega^2}\,(a^2-l^2)}}\,.
  \label{keg}
\end{align}

The fully explicit form of the metric \eqref{newmetricGP2005-0} is thus quite complicated because substituting  \eqref{a3}--\eqref{keg} into \eqref{P} and \eqref{Q} gives cumbersome expressions. Another fundamental problem is the actual physical meaning of the 7 parameters ${\tilde{m} ,a, l, \tilde{e}, \tilde{g}, \alpha, \tilde{\Lambda}}$. These have been clearly interpreted only in special subcases when some of the other parameters were set to zero. In such subcases, they represent \emph{mass, Kerr-like rotation, NUT parameter, electric charge, magnetic charge, acceleration}, and \emph{cosmological constant}, respectively. Their meaning in a \emph{completely general situation} is still an open problem. Moreover, there is an additional (auxiliary) \emph{twist parameter} $\omega$. In previous works  \cite{GriffithsPodolsky:2005,GriffithsPodolsky:2006,PodolskyGriffiths:2006} it was argued that $\omega$ is related \emph{both} to $a$ and $l$, and in some cases can be scaled appropriately using the remaining coordinate freedom. A satisfactory insight into all these problems is still missing. It is the aim of the present work to clarify such issues. We achieve this by presenting a new compact, explicit and considerably simplified form of the Pleba\'nski--Demia\'nski metric, namely \eqref{newmetricGP2005}--\eqref{finalQagain}, for a complete family of black holes.

The first step in improving the form of the spacetime is to introduce a \emph{new set of the mass and charge parameters} $m, e, g$. Following our previous paper \cite{PodolskyVratny:2021}, we define them as
\begin{eqnarray}
m   \equi S\,\tilde{m} - \alpha \frac{l}{\omega} (a^2-l^2  + e^2 + g^2)\,, \nonumber\\
e^2 \equi S\, \tilde{e}^2 \,, \label{PD-par-trans}\\[2mm]
g^2 \equi S\, \tilde{g}^2 \,, \nonumber
\end{eqnarray}
where $S$ is a \emph{specific scaling constant}
\begin{equation}
  S \equiv \frac{a^2-l^2}{\omega^2 k} \,.
  \label{Sdef}
\end{equation}
Notice that
\begin{equation}
  (\omega^2 k+\tilde{e}^2+\tilde{g}^2) = S^{-1}\,(a^2-l^2+e^2+g^2) \,,
  \label{aux1}
\end{equation}
which is a much simpler expression than \eqref{keg}.

In terms of these new parameters $m, e, g$, the coefficients \eqref{a3}--\eqref{n} take the form
\begin{align}
a_3 &=  S^{-1}\,\frac{a}{\omega}\,
\Big[\,2\alpha m - 2\alpha^2 \frac{l}{\omega} (a^2-l^2  + e^2 + g^2)
    -\frac{4}{3}\tilde{\Lambda}S\,l\,\omega\,\Big]\,,  \label{a3new}\\
a_4 &= -S^{-1} \,\frac{a^2}{\omega^2}\,\Big[\,\alpha^2 \,(a^2-l^2+e^2+g^2) +\frac{1}{3}\tilde{\Lambda}S\,\omega^2\,\Big]\,, \label{a4new}\\
\epsilon &= S^{-1}\, \Big[\,1 + 4\alpha\,\frac{l}{\omega}\,m
   -\alpha^2\,\frac{a^2-l^2}{\omega^2}\,(a^2-l^2+e^2+g^2)
   -\frac{1}{3}\tilde{\Lambda}S\,(a^2+3l^2)\, \Big],
  \label{epsilonnew}\\
n &= S^{-1}\, \Big[\,l -\alpha\,\frac{a^2-l^2}{\omega}\,m  + \frac{1}{3}\tilde{\Lambda}S\,(a^2-l^2)\,l\,\Big].
  \label{nnew}
\end{align}
The key metric functions \eqref{P}, \eqref{Q} thus nicely simplify to
\begin{align}
\PP(\theta) = S^{-1}\, P(\theta)\,, \qquad\qquad
\QQ(r)      = S^{-1}\, Q (r)\,, \label{QP relations}
\end{align}
where
\begin{align}
P(\theta)&= 1
   -2\,\Big(\,\frac{\alpha}{\omega}\,m - \frac{1}{3}\tilde{\Lambda}S\,l \Big)(l+a\,\cos\theta)\nonumber\\
      &\qquad  +\Big(\,\frac{\alpha^2}{\omega^2} (a^2-l^2  + e^2 + g^2) + \frac{1}{3}\tilde{\Lambda}S \Big)(l+a\,\cos\theta)^2 \,,
\label{defnewP-omega}\\
Q (r) &= \Big[\,r^2 - 2m\, r  + (a^2-l^2+e^2+g^2) \,\Big]
         \Big(1+\alpha\,\frac{a-l}{\omega}\,r\Big) \Big(1-\alpha\,\frac{a+l}{\omega}\,r\Big) \nonumber\\
      &\qquad  - \frac{1}{3}\tilde{\Lambda}S\,r^2 \Big[\,r^2 + 2\alpha\,\frac{l}{\omega}\,(a^2-l^2)\,r + (a^2+3l^2)\,\Big]. \label{defnewQ-omega}
\end{align}

With \eqref{QP relations}, the metric \eqref{newmetricGP2005-0} now reads
\begin{align}
\dd \tilde{s}^2 = \frac{S}{\Omega^2} &
  \bigg(\!-\frac{Q}{\rho^2}\,S^{-2}\big[\dd t- \big(a\sin^2\theta +4l\sin^2\!{\textstyle\frac{1}{2}\theta} \big)\dd\varphi \big]^2 + \frac{\rho^2}{Q}\,\dd r^2  \nonumber\\
& \quad  + \frac{\rho^2}{P}\,\dd\theta^2
  + \frac{P}{\rho^2}\,\sin^2\theta\, S^{-2}\big[ a\dd t -\big(r^2+(a+l)^2\big)\,\dd\varphi \big]^2
 \bigg). \label{transfmetric}
\end{align}
Recall that it is a solution to the Einstein--Maxwell field equations with a cosmological constant~$\tilde{\Lambda}$.

As the second step, we now \emph{rescale the coordinates~$t$ and~$\varphi$} by a \emph{constant scaling factor} ${S \ne 0}$. (This is possible because their ranges have not yet been specified.) In other words, we perform the transformation
\begin{align}
t \to S\,t\,, \qquad\qquad
\varphi \to S\,\varphi\,, \label{rescaling}
\end{align}
which completely removes all the constants $S$ from the conformally related metric
\begin{align}
 \dd s^2 \equiv S^{-1}\,\dd \tilde{s}^2 \,, \label{rescaledmetric}
\end{align}
that is
\begin{align}
\dd s^2 = \frac{1}{\Omega^2} &
  \bigg(\!-\frac{Q}{\rho^2}\,\big[\dd t- \big(a\sin^2\theta +4l\sin^2\!{\textstyle\frac{1}{2}\theta} \big)\dd\varphi \big]^2 + \frac{\rho^2}{Q}\,\dd r^2  \nonumber\\
& \quad  + \frac{\rho^2}{P}\,\dd\theta^2
  + \frac{P}{\rho^2}\,\sin^2\theta\, \big[ a\dd t -\big(r^2+(a+l)^2\big)\,\dd\varphi \big]^2
 \bigg). \label{newmetric}
\end{align}

Since the energy--momentum tensor of the Maxwell field ${4\pi\,T_{ab} = F_{ac} {F_{b}}^{c} - \tfrac{1}{4} g_{ab} F_{cd}F^{cd}}$ in four dimensions is trace-free, Einstein's equations read ${R_{ab}=\Lambda\,g_{ab}+8\pi\,T_{ab} }$, and the Ricci scalar is  ${R=4\Lambda}$. Under the constant conformal rescaling \eqref{rescaledmetric} of the metric, the Ricci tensor is invariant: ${ g_{ab} = S^{-1}\, \tilde{g}_{ab}}$ implies ${R_{ab} = \tilde{R}_{ab}}$ and ${ R = \tilde{R}\, S}$. Consequently, the new metric \eqref{newmetric} is a solution to the Einstein--Maxwell field equations with a cosmological constant~$\Lambda$, provided
\begin{align}
 \Lambda \equiv \tilde{\Lambda}\, S \,,\qquad\qquad
 F_{ab} \equiv  \tilde{F}_{ab}\,\sqrt{S} \,.
  \label{rescaledLambda}
\end{align}
The corresponding metric functions \eqref{defnewP-omega}, \eqref{defnewQ-omega} are thus
 \begin{align}
P(\theta)&= 1
   -2\,\Big(\,\frac{\alpha}{\omega}\,m - \frac{\Lambda}{3}\,l \Big)(l+a\,\cos\theta)\nonumber\\
      &\qquad  +\Big(\,\frac{\alpha^2}{\omega^2} (a^2-l^2  + e^2 + g^2) + \frac{\Lambda}{3} \Big)(l+a\,\cos\theta)^2 \,,
\label{defnewP}\\
Q (r) &= \Big[\,r^2 - 2m\, r  + (a^2-l^2+e^2+g^2) \Big]
         \Big(1+\alpha\,\frac{a-l}{\omega}\,r\Big) \Big(1-\alpha\,\frac{a+l}{\omega}\,r\Big) \nonumber\\
      &\qquad  - \frac{\Lambda}{3}\,r^2 \Big[\,r^2 + 2\alpha\,\frac{l}{\omega}\,(a^2-l^2)\,r + (a^2+3l^2)\,\Big]. \label{defnewQ}
\end{align}

As the third step, it remains to \emph{fix the auxiliary twist parameter $\omega$}, coupled with \emph{both} the Kerr-like rotation~$a$ \emph{and} the NUT parameter~$l$. It was found in \cite{Vratny:2018} and conveniently employed in \cite{MatejovPodolsky:2021, MatejovPodolsky:2022, PodolskyVratny:2021} that the most suitable gauge choice of this twist parameter is
\begin{align}
\omega \equiv \frac{a^2 + l^2}{a}\,, \label{ChoiceOmega}
\end{align}
so that
\begin{align}
\frac{a}{\omega} = \frac{a^2}{a^2 + l^2}\,, \qquad\qquad \frac{l}{\omega} = \frac{a\,l}{a^2 + l^2}\,. \label{ChoiceOmega-al}
\end{align}
Substituting these expressions into \eqref{Omega}, \eqref{defnewP} and \eqref{defnewQ}, we obtain the explicit functions $\Omega$, $P$ and~$Q$, namely
\begin{eqnarray}
\Omega    \rovno 1-\frac{\alpha\,a}{a^2+l^2}\,r\,(l+a \cos \theta) \,, \label{finalOmega}\\
P(\theta) \rovno 1
   -2\,\Big(\,\frac{\alpha\,a}{a^2+l^2}\,\,m - \frac{\Lambda}{3}\,l \Big)(l+a\,\cos\theta)\nonumber\\
      &&  +\Big(\frac{\alpha^2 a^2}{(a^2+l^2)^2} (a^2-l^2 + e^2 + g^2) + \frac{\Lambda}{3} \Big)(l+a\,\cos\theta)^2 \,,
 \label{finalP}\\
Q(r) \rovno \Big[\,r^2 - 2m\, r  + (a^2-l^2+e^2+g^2) \Big]
            \Big(1+\alpha\,a\,\frac{a-l}{a^2+l^2}\, r\Big)
            \Big(1-\alpha\,a\,\frac{a+l}{a^2+l^2}\, r\Big)\nonumber\\
   &&  - \frac{\Lambda}{3}\,r^2 \Big[\,r^2 + 2\alpha\,a\,l\,\frac{a^2-l^2}{a^2+l^2}\,r + (a^2+3l^2)\,\Big]. \label{finalQ}
\end{eqnarray}

In fact, for a generic class of black holes the metric functions $P$ and $Q$ can be further simplified.  To this end, let us define convenient parameters $\mu$, $\lambda$, and $\A$ (representing the ``modified'' mass, cosmological constant, and acceleration, respectively) as
\begin{align}
\mu     & \equiv  m - \lambda\,\A \qquad = m - \frac{\Lambda}{3}\,l\,\frac{a^2+l^2}{\alpha\,a}\,,  \label{def-mu}\\[2mm]
\lambda & \equiv \frac{\Lambda}{3}\,\frac{(a^2+l^2)^2}{\alpha^2 a^2} \,, \label{def-lambda}\\[2mm]
\A      & \equiv \frac{\alpha\,a\,l}{a^2+l^2} \,.  \label{def-cal-A}
\end{align}
Moreover, we introduce a \emph{pair of special constants} $r_{\Lambda+}$ and $r_{\Lambda-}$ by
\begin{align}
r_{\Lambda \pm} &\equiv \mu \pm \sqrt{\mu^2 + l^2 - a^2 - e^2 - g^2 - \lambda} \,. \label{rLambda+-}
\end{align}

From these definitions it immediately follows that
\begin{align}
\frac{\alpha\,a}{a^2+l^2}\,(r_{\Lambda+}+r_{\Lambda-})&= 2\Big(\,\frac{\alpha\,a}{a^2+l^2}\,m - \frac{\Lambda}{3}\,l\, \Big)\,, \nonumber\\
\frac{\alpha^2 a^2}{(a^2+l^2)^2}\,r_{\Lambda+}\,r_{\Lambda-} &= \frac{\alpha^2 a^2}{(a^2+l^2)^2} (a^2-l^2 + e^2 + g^2) + \frac{\Lambda}{3}\,, \label{rLambda+-2}
\end{align}
so that \eqref{finalP} can be re-expressed as
\begin{align}
P(\theta) = \Big(1 -\frac{\alpha\,a}{a^2+l^2}\,r_{\Lambda+}\,(l+a\,\cos\theta) \Big)
            \Big(1 -\frac{\alpha\,a}{a^2+l^2}\,r_{\Lambda-}\,(l+a\,\cos\theta) \Big) \,.
 \label{factorizedP}
\end{align}
The metric function $P(\theta)$ \emph{is thus nicely factorized}.

Using \eqref{def-mu}--\eqref{rLambda+-}, the expression \eqref{finalQ} for the metric function $Q(r)$ is also simplified to
\begin{align}
Q(r) = &\Big[\,r^2 - 2\mu\, r  + (a^2-l^2+e^2+g^2+\lambda) \Big]
            \Big(1+\alpha\,a\,\frac{a-l}{a^2+l^2}\, r\Big)
            \Big(1-\alpha\,a\,\frac{a+l}{a^2+l^2}\, r\Big)\nonumber\\
   &- \lambda\, \Big[\,1+ \frac{\alpha^2 a^2}{(a^2+l^2)^2}\,r^4\,\Big].
 \label{simplerQ}
\end{align}
In the cases when ${\mu^2+l^2>a^2+e^2+g^2+\lambda}$, the definition \eqref{rLambda+-} yields \emph{two real distinct constants} $r_{\Lambda+}$ and $r_{\Lambda-}$, and \eqref{simplerQ} takes the form
\begin{align}
Q(r) = &\big(r-r_{\Lambda+} \big) \big( r-r_{\Lambda-} \big)
            \Big(1+\alpha\,a\,\frac{a-l}{a^2+l^2}\, r\Big)
            \Big(1-\alpha\,a\,\frac{a+l}{a^2+l^2}\, r\Big) - \frac{\Lambda}{3}\,\Big[\,r^4 + \frac{(a^2+l^2)^2}{\alpha^2 a^2}\,\Big].
 \label{factorQ}
\end{align}

Interestingly, \emph{when} ${\Lambda=0}$, the constants $r_{\Lambda \pm}$ defined by \eqref{rLambda+-} reduce to
\begin{eqnarray}
r_{\pm} \equi  m  \pm \sqrt{\,m^2+l^2-a^2-e^2-g^2}\,. \label{r+-}
\end{eqnarray}
These parameters then identify (independently of the acceleration $\alpha$) the \emph{two black-hole horizons} because they are also the roots of the metric functions $Q(r)$ given by \eqref{factorQ}, cf. \cite{PodolskyVratny:2021}.

Finally, although the unique scaling constant $S$ defined by \eqref{Sdef} \emph{does not enter the final form} of the metric \eqref{newmetric} with \eqref{finalOmega}--\eqref{finalQ}, it may be useful to present its explicit form in terms of the new parameters. Substitution from \eqref{PD-par-trans}  into \eqref{kfrac} with ${\Lambda = \tilde{\Lambda}\, S }$ yields the relation
\begin{equation}
  S = 1 - 2\alpha \frac{l}{\omega}\, m  + \alpha^2 \frac{l^2}{\omega^2}\,(a^2-l^2+e^2+g^2) + \Lambda\, l^2 \,,  \label{S}
\end{equation}
that is, using \eqref{ChoiceOmega}, \eqref{def-cal-A}--\eqref{rLambda+-2},
\begin{equation}
  S  = (1-\A\,r_{\Lambda+})(1-\A\,r_{\Lambda-}) \,.  \label{Salter}
\end{equation}
The rescaling transformation (\ref{rescaledmetric}) thus actually removes two coordinate singularities hidden in the expression (\ref{Salter})  at ${\A\,r_{\Lambda\pm} = 1}$. This fact was already observed for the ${\Lambda=0}$ case in our previous article \cite{PodolskyVratny:2021}.

Moreover, it can be seen that ${S=1}$ whenever ${\A\,r_{\Lambda+} = 0 = \A\,r_{\Lambda-}}$. For ${\Lambda=0}$, this happens if ${l=0}$ or ${\alpha=0}$ or ${a=0}$, in which cases ${m=\tilde{m}}$, ${e=\tilde{e}}$, ${g=\tilde{g}}$.

\newpage

For ${\Lambda\ne0}$, the value of the scaling factor is generically ${S\ne1}$. In the case ${l=0}$ it follows from \eqref{S} that ${S=1}$, but in the case ${l\ne0}$ we get ${S = 1 + \Lambda\, l^2}$ even if ${\alpha=0}$ or ${a=0}$. Generally, ${S=1}$ \emph{only for a special value} of the cosmological constant
\begin{eqnarray}
\Lambda = \frac{\alpha \, a}{a^2+l^2} \,\Big[\, 2\,\frac{m}{l}- \frac{\alpha \, a}{a^2+l^2}\big( a^2-l^2+e^2+g^2 \big) \Big]\,.
\label{S=1condition}
\end{eqnarray}

\section{Summary of the new form of a generic black hole}
\label{sec:introsummary}

It is now useful to summarize our new metric representation of the \emph{complete family of black holes} contained in the class of Pleba\'nski--Demia\'nski spacetimes \cite{PlebanskiDemianski:1976}. Recall that such spacetimes are the most general exact solutions to Einstein--Maxwell equations of algebraic type~D with double-aligned non-null electromagnetic field (see Chapter~16 of the monograph \cite{GriffithsPodolsky:2009} for the recent review and number of related references).

The new metric form, which improves the previous representation found by Griffiths and Podolsk\'y \cite{GriffithsPodolsky:2005,GriffithsPodolsky:2006,PodolskyGriffiths:2006}, reads
\begin{align}
\dd s^2 = \frac{1}{\Omega^2} &
  \bigg(\!-\frac{Q}{\rho^2}\big[\,\dd t- \big(a\sin^2\theta +4l\sin^2\!\tfrac{1}{2}\theta \big)\dd\varphi \big]^2
   + \frac{\rho^2}{Q}\,\dd r^2  \nonumber\\
& \quad  + \,\frac{\rho^2}{P}\,\dd\theta^2
  + \frac{P}{\rho^2}\,\sin^2\theta\, \big[ a\,\dd t -\big(r^2+(a+l)^2\big)\,\dd\varphi \big]^2
 \bigg), \label{newmetricGP2005}
\end{align}
where
\begin{eqnarray}
\Omega    \rovno 1-\frac{\alpha\,a}{a^2+l^2}\, r\,(l+a \cos \theta) \,, \label{newOmega}\\[2mm]
\rho^2    \rovno r^2+(l+a \cos \theta)^2 \,, \label{newrho}\\[2mm]
P(\theta) \rovno 1
   -2\,\Big(\,\frac{\alpha\,a}{a^2+l^2}\,\,m - \frac{\Lambda}{3}\,l \Big)(l+a\,\cos\theta)\nonumber\\
      &&  +\Big(\frac{\alpha^2 a^2}{(a^2+l^2)^2} (a^2-l^2 + e^2 + g^2) + \frac{\Lambda}{3} \Big)(l+a\,\cos\theta)^2 \,,
 \label{finalPagain}\\[2mm]
Q(r) \rovno \Big[\,r^2 - 2m\, r  + (a^2-l^2+e^2+g^2) \Big]
            \Big(1+\alpha\,a\,\frac{a-l}{a^2+l^2}\, r\Big)
            \Big(1-\alpha\,a\,\frac{a+l}{a^2+l^2}\, r\Big)\nonumber\\
   &&  - \frac{\Lambda}{3}\,r^2 \Big[\,r^2 + 2\alpha\,a\,l\,\frac{a^2-l^2}{a^2+l^2}\,r + (a^2+3l^2)\,\Big]. \label{finalQagain}
\end{eqnarray}

The spacetime depends on \emph{seven physical parameters}, namely
\begin{align}
m \quad.....\quad & \hbox{mass parameter}     \nonumber \,,\\
a \quad.....\quad & \hbox{Kerr-like rotation} \nonumber \,,\\
l \quad.....\quad & \hbox{NUT parameter}      \nonumber \,,\\
e \quad.....\quad & \hbox{electric charge}    \nonumber \,,\\
g \quad.....\quad & \hbox{magnetic charge}    \nonumber \,,\\
\alpha \quad.....\quad & \hbox{acceleration}  \nonumber \,,\\
\Lambda\quad.....\quad & \hbox{cosmological constant}  \nonumber \,.
\end{align}

This metric is  compact and fully explicit, and the ambiguous twist parameter $\omega$ has been removed by its most convenient choice. Moreover, the standard forms of famous black hole spacetimes --- namely Kerr--Newman--(A)dS, charged Taub--NUT--(A)dS, their accelerated versions, and others --- can easily be obtained as direct subcases of \eqref{newmetricGP2005}--\eqref{finalQagain} by setting the corresponding physical parameters to zero.

\newpage

When ${\Lambda=0}$, both metric functions $P$ and $Q$ are factorized, see \cite{PodolskyVratny:2021} for more details. With ${\Lambda\not=0}$ this cannot be in general achieved. However, it is possible to explicitly factorize the function $P$ and compactify the function $Q$ as
\begin{eqnarray}
P(\theta) \rovno \Big(1 -\frac{\alpha\,a}{a^2+l^2}\,r_{\Lambda+}\,(l+a\,\cos\theta) \Big)
            \Big(1 -\frac{\alpha\,a}{a^2+l^2}\,r_{\Lambda-}\,(l+a\,\cos\theta) \Big) , \label{newP}\\[1mm]
Q(r) \rovno \big(r-r_{\Lambda+} \big) \big( r-r_{\Lambda-} \big)
            \Big(1+\alpha\,a\,\frac{a-l}{a^2+l^2}\, r\Big)
            \Big(1-\alpha\,a\,\frac{a+l}{a^2+l^2}\, r\Big) - \frac{\Lambda}{3}\,\Big[\,r^4 + \frac{(a^2+l^2)^2}{\alpha^2 a^2}\,\Big],\qquad \label{newQ}
\end{eqnarray}
using the two specific constants
\begin{eqnarray}
r_{\Lambda \pm} \equi  \mu \pm \sqrt{\mu^2 + l^2 - a^2 - e^2 - g^2 - \lambda}\,, \label{r+}
\end{eqnarray}
where
\begin{align}
\mu     \equiv   m - \frac{\Lambda}{3}\,l\,\frac{a^2+l^2}{\alpha\,a}\,,  \qquad
\lambda \equiv \frac{\Lambda}{3}\,\frac{(a^2+l^2)^2}{\alpha^2 a^2} \,. \label{def-mu-and-lambda}
\end{align}
This is possible provided ${\mu^2 + l^2 > a^2 + e^2 + g^2 + \lambda}$, in which case the expressions \eqref{r+} yield two distinct real constants (or a double root of $P$ given by ${r_{\Lambda+}  = r_{\Lambda-}= \mu}$ in the specific situation when ${\mu^2 + l^2 = a^2 + e^2 + g^2 + \lambda}$).

The new form of the metric \eqref{newmetricGP2005}--\eqref{finalQagain} nicely represents the \emph{complete family of type D black holes}. Moreover, it naturally generalizes the standard forms of the most important black hole solutions, with two black-hole horizons (outer and inner) and two cosmological/acceleration horizons.

\section{The main subclasses of type~D black holes}
\label{sec_subclasses}

These are easily obtained  by setting the appropriate physical parameters to zero, as follows.

\subsection{Black holes in flat universe\\ (${\Lambda=0}$\,: no cosmological constant)}
\label{subclasse-Lambda=0}

In the case ${\Lambda=0}$, we get ${\mu=m}$ and ${\lambda=0}$. When ${m^2 + l^2 > a^2 + e^2 + g^2}$ (which guarantees that two distinct roots $r_+$ and $r_-$ exist) the metric functions \eqref{newP}, \eqref{newQ} thus take the form
\begin{align}
P(\theta)&= \Big( 1-\frac{\alpha\,a}{a^2+l^2}\, r_{+} (l+a \cos \theta) \Big)
            \Big( 1-\frac{\alpha\,a}{a^2+l^2}\, r_{-} (l+a \cos \theta) \Big)\,,
\label{newPfactorized}\\
Q (r) &= \big(r-r_{+} \big) \big( r-r_{-} \big)
         \Big(1+\alpha\,a\,\frac{a-l}{a^2+l^2}\, r\Big)
         \Big(1-\alpha\,a\,\frac{a+l}{a^2+l^2}\, r\Big), \label{newQsemifactorized}
\end{align}
where
\begin{eqnarray}
r_{\pm} \equi  m  \pm \sqrt{\,m^2+l^2-a^2-e^2-g^2}\,, \label{r+-repeated}
\end{eqnarray}
cf. \eqref{r+-}. The constants $r_+$ and $r_-$ now directly identify (independently of the acceleration $\alpha$) the \emph{two black-hole horizons} because they are also the \emph{roots of the metric functions} $Q(r)$ given by \eqref{newQsemifactorized}. This large family of black holes was thoroughly analyzed in our previous work \cite{PodolskyVratny:2021}, and it is not necessary to repeat all the arguments and results here.

\subsection{Kerr--Newman--NUT--(anti-)de Sitter black holes\\
 (${\alpha=0}$\,: no acceleration)}

By setting the acceleration parameter $\alpha$ to zero, the metric function \eqref{newOmega} reduces to
${\Omega=1}$, while \eqref{newrho} remains the same. Concerning the functions $P$ and $Q$ given by \eqref{newP} and \eqref{newQ}, respectively, one has to be more careful in evaluating the limits of the terms ${\alpha\,a\, r_{\Lambda\pm}}$ because the acceleration ${\alpha\to0}$ appears also in the denominator of the parameters $\mu$ and $\lambda$, defined by \eqref{def-mu-and-lambda}, which enter $r_{\Lambda\pm}$. In this case it is more convenient to directly set ${\alpha=0}$ in the most general forms of these metric functions  \eqref{finalPagain} and \eqref{finalQagain}. In any case, we obtain the metric
\begin{align}
\dd s^2 = &
  -\frac{Q}{\rho^2}\left[\,\dd t- \left(a\sin^2\theta +4l\sin^2\!{\textstyle\frac{1}{2}\theta} \right)\dd\varphi \right]^2 + \frac{\rho^2}{Q}\,\dd r^2  \nonumber\\
& \quad  + \,\frac{\rho^2}{P}\,\dd\theta^2
  + \frac{P}{\rho^2}\,\sin^2\theta\, \big[ a\,\dd t -\big(r^2+(a+l)^2\big)\,\dd\varphi \big]^2
 \,, \label{metric-alpha=0}
\end{align}
where
\begin{eqnarray}
\rho^2 \rovno r^2+(l+a \cos \theta)^2 \,. \label{rho-alpha=0}\\
P(\theta) \rovno 1
   + 2\,\frac{\Lambda}{3}\,l\,(l+a\,\cos\theta) + \frac{\Lambda}{3} \,(l+a\,\cos\theta)^2 \,,
 \label{P-alpha=0}\\
Q(r) \rovno r^2 - 2m\, r  + (a^2-l^2+e^2+g^2)
            - \frac{\Lambda}{3}\,r^2 \big(\,r^2 + a^2+3l^2 \big)\,. \label{Q-alpha=0}
\end{eqnarray}

This result is the same as the limit ${\alpha\to0}$  of the metric functions \eqref{newP} and \eqref{newQ}. Indeed,
\begin{align}
 & \lim_{\alpha\to0 }\,\frac{\alpha\,a}{a^2+l^2}\,r_{\Lambda\pm}
   = - \frac{\Lambda}{3}\,l \pm \sqrt{\Big(\frac{\Lambda}{3}\, l \Big)^2 - \frac{\Lambda}{3}}
   \equiv L_\pm \,, \label{L+-}\\[2mm]
 & \lim_{\alpha\to0 }\, \A\,r_{\Lambda\pm} = l\,L_\pm \,,
   \label{limA+-}
\end{align}
so that
\begin{align}
L_+ + L_- =  - 2\,\frac{\Lambda}{3}\,l\,, \qquad\qquad  L_+ \,L_- =   \frac{\Lambda}{3} \,.   \label{L+L-}
\end{align}
Thus ${\displaystyle\lim_{\alpha\to0 }\,P(\theta) = \big(1 - L_+\,(l+a\,\cos\theta) \big) \big(1 - L_-\,(l+a\,\cos\theta) \big)}$ gives \eqref{P-alpha=0}, which can be rewritten~as
\begin{align}
P(\theta) = (1 + \Lambda\, l^2) + \frac{4}{3}\,\Lambda\,a\,l\,\cos\theta + \frac{1}{3}\,\Lambda \,a^2 \cos^2 \theta \,.
 \label{P-alpha=0-alternative}
\end{align}
In a similar way, the limit of \eqref{newQ} using \eqref{rLambda+-2} yields \eqref{Q-alpha=0}.
Moreover, in the limit of vanishing acceleration the scaling factor \eqref{Salter}, using \eqref{limA+-} and \eqref{L+L-}, becomes
\begin{align}
\lim_{\alpha\to0 }\,S   = 1 + \Lambda\, l^2 \,.   \label{lim-of-S-alpha->0}
\end{align}

We must emphasize that the forms \eqref{P-alpha=0-alternative} and \eqref{Q-alpha=0} of the metric functions $P(\theta)$ and $Q(r)$ \emph{are different} from the analogous metric functions for the Kerr--Newman--NUT--(anti-)de Sitter black holes as given by Eq.~(16.23) in \cite{GriffithsPodolsky:2009}. In fact, they are \emph{equivalent re-parametrization} of this solution. Indeed, we have to take into account the nontrivial scaling \eqref{QP relations}, that is
\begin{align}
\PP(\theta) = S^{-1}\, P(\theta)\,, \qquad\qquad
\QQ(r)      = S^{-1}\, Q (r)\,, \label{QP relations-repeated}
\end{align}
where $S$ is the constant \eqref{lim-of-S-alpha->0}. Straightforward calculation using the relations
\eqref{PD-par-trans}, \eqref{rescaledLambda} between the physical parameters then yields
\begin{eqnarray}
\PP(\theta) \rovno 1 + \tfrac{4}{3}\tilde{\Lambda}\,a\,l\,\cos\theta + \tfrac{1}{3}\tilde{\Lambda} \,a^2 \cos^2 \theta \,,
 \label{P-alpha=0-BOOK}\\[3pt]
\QQ(r) \rovno (a^2-l^2+\tilde{e}^2+{\tilde g}^2) - 2\tilde{m}\, r  + r^2
            - \tilde{\Lambda}\, \Big[ (a^2-l^2)\,l^2 + \big(\,\tfrac{1}{3} a^2+2l^2 \big)\,r^2 + \tfrac{1}{3}\,r^4\,\Big]\,, \label{Q-alpha=0-BOOK}
\end{eqnarray}
which is exactly the form of the metric functions given by Eq.~(16.23) in \cite{GriffithsPodolsky:2009}.

All famous subcases of this general family of (non-accelerating) Kerr--Newman--NUT--(anti-)de Sitter black holes, expressed now in a compact way by the metric \eqref{metric-alpha=0} with \eqref{rho-alpha=0}--\eqref{Q-alpha=0} [or  \eqref{P-alpha=0-alternative}, equivalent to \eqref{P-alpha=0}], are readily obtained. These are the black hole solutions of Kerr--Newman--(anti-)de Sitter (${l=0}$), charged Taub--NUT--(anti-)de Sitter (${a=0}$), Kerr--(anti-)de Sitter (${l=0}$, ${e=0=g}$), Reissner--Nordstr\"{o}m--(anti-)de Sitter (${a=0}$, ${l=0}$), and Schwarzschild--(anti-)de Sitter (${a=0}$, ${l=0}$, and ${e=0=g}$). Of course, by setting ${\Lambda=0}$, the corresponding black holes in asymptotically flat universe are obtained (the same as in Sec.~\ref{subclasse-Lambda=0}).

\subsection{Accelerating Kerr--Newman--(anti-)de Sitter black holes\\
 (${l=0}$\,: no NUT)}

Without the NUT parameter $l$, the new metric \eqref{newmetricGP2005} reduces to
\begin{align}
\dd s^2 = \frac{1}{\Omega^2} &
  \left(-\frac{Q}{\rho^2}\left[\,\dd t - a\sin^2\theta\, \dd\varphi \right]^2 + \frac{\rho^2}{Q}\,\dd r^2 \right. \nonumber\\
& \quad \left. + \,\frac{\rho^2}{P}\,\dd\theta^2
  + \frac{P}{\rho^2}\,\sin^2\theta\, \big[ a\,\dd t - (r^2+a^2)\,\dd\varphi \big]^2
 \right), \label{metric-l=0}
\end{align}
where
\begin{eqnarray}
\Omega    \rovno 1-\alpha\, r\,\cos\theta \,, \label{Omega-l=0}\\[1mm]
\rho^2    \rovno r^2+a^2\cos ^2\theta \,, \label{rho-l=0}\\[2mm]
P(\theta) \rovno \big( 1-\alpha\,r_{\Lambda+} \cos\theta \big)
                 \big( 1-\alpha\,r_{\Lambda-} \cos\theta \big) , \label{P-l=0}\\
Q(r) \rovno \big(r-r_{\Lambda+} \big) \big( r-r_{\Lambda-} \big)
            \big(1+\alpha\,r\big)
            \big(1-\alpha\,r\big) - \frac{\Lambda}{3}\,\Big(r^4 + \frac{a^2}{\alpha^2}\Big). \label{Q-l=0}
\end{eqnarray}
where the specific constants $r_{\Lambda\pm}$ are now simplified to
\begin{eqnarray}
r_{\Lambda\pm} \equi  m \pm \sqrt{ m^2 - a^2 - e^2 - g^2 - \frac{\Lambda}{3}\,\frac{a^2}{\alpha^2}}\,. \label{r+-l=0}
\end{eqnarray}
The metric functions $P(\theta)$ and $Q(r)$ can be equivalently rewritten as
\begin{eqnarray}
P(\theta) \rovno 1
   -2\,\alpha\,m\,\cos\theta +\Big[\,\alpha^2 (a^2 + e^2 + g^2) + \frac{\Lambda}{3}\,a^2 \,\Big] \cos^2\theta \,,
 \label{finalPl=0}\\
Q(r) \rovno \big[\,r^2 - 2m\, r  + (a^2+e^2+g^2) \big]
            \big(1+\alpha\, r\big)
            \big(1-\alpha\, r\big) - \frac{\Lambda}{3}\,r^2 \big[\,r^2 + a^2 \,\big]. \label{finalQl=0}
\end{eqnarray}

In this explicit form we easily obtain all possible subcases by simply setting the corresponding physical parameters to zero. For vanishing acceleration (${\alpha=0}$), the metric of the Kerr--Newman--(anti-)de Sitter black hole solution is recovered, which then yields the standard form of the Kerr--Newman solution in the Boyer--Lindquist coordinates in the case of vanishing cosmological constant (${\Lambda=0}$). Contrarily, by setting ${\Lambda=0}$ first, we obtain the general metric of accelerating Kerr--Newman black holes. For vanishing charges (${e=0=g}$), it is equivalent to the rotating $C$-metric, first identified by Hong and Teo \cite{HongTeo:2005}.

\subsection{Charged Taub--NUT--(anti-)de Sitter black holes\\
 (${a=0}$\,: no rotation)}
 \label{norotation}

By setting the Kerr-like rotation parameter $a$ to zero, the new metric \eqref{newmetricGP2005} considerably simplifies and \emph{becomes independent of the acceleration}~$\alpha$ (because the metric functions \eqref{newOmega}--\eqref{newQ} depend on $\alpha$ only via the product ${\alpha\,a}$). Indeed, ${\Omega=1}$ and
${P = 1 + \Lambda\,l^2}$,  so that
\begin{align}
\dd s^2 = &
  -\frac{Q}{\rho^2} \big(\dd t - 4l\sin^2\!{\textstyle\frac{1}{2}\theta} \,\dd\varphi \big)^2 + \frac{\rho^2}{Q}\,\dd r^2
  + \rho^2\,\Big(\,\frac{\dd\theta^2}{1 + \Lambda\,l^2} + (1 + \Lambda\,l^2) \sin^2\theta\, \dd\varphi^2 \Big)\,, \label{metric-a=0}
\end{align}
where
\begin{eqnarray}
Q(r) \rovno (1 - \Lambda\,l^2)\,r^2 - 2m\, r  + (e^2+g^2-l^2) - \frac{\Lambda}{3}\,r^4   \,, \label{rho-a=0}\\
\rho^2  \rovno r^2+l^2  \,. \label{Q-a=0}
\end{eqnarray}
This explicitly demonstrates that \emph{there is no accelerating ``purely'' NUT--(anti-)de Sitter black hole in the Pleba\'nski--Demia\'nski family} of spacetimes.

For ${\Lambda=0}$, this observation was made already in the original works \cite{GriffithsPodolsky:2005,GriffithsPodolsky:2006,PodolskyGriffiths:2006}, and recently clarified in \cite{PodolskyVratny:2020}. It was proven that the metric for accelerating (non-rotating) black holes with purely NUT parameter --- which was found by  Chng, Mann and Stelea \cite{ChngMannStelea:2006} in 2006  and analyzed in detail in \cite{PodolskyVratny:2020} --- is of algebraic type~I. Therefore, it \emph{cannot} be contained in the Pleba\'nski--Demia\'nski class which is of type~D. We have just shown that the same is true also in the case of a non-vanishing cosmological constant $\Lambda$.

It should again be emphasized that the metric function \eqref{rho-a=0} for $Q(r)$ \emph{is different} from the analogous metric function for the charged Taub--NUT--(anti-)de Sitter black hole as given by Eq.~(12.19) in \cite{GriffithsPodolsky:2009}. Actually, it is simpler. Such a difference is caused by the nontrivial rescaling ${S = 1 + \Lambda\, l^2 }$, see \eqref{lim-of-S-alpha->0}, \eqref{QP relations-repeated}. Considering the relations \eqref{PD-par-trans}, \eqref{QP relations} and \eqref{rescaledLambda}, we get
\begin{eqnarray}
\PP(\theta) \rovno 1 \,, \label{P-a=0-BOOK}\\[3pt]
\QQ(r) \rovno  r^2 -l^2 - 2\tilde{m}\, r +\tilde{e}^2+{\tilde g}^2
            - \tilde{\Lambda}\, \big( \tfrac{1}{3}\,r^4 + 2l^2\,r^2 -l^4 \big)\,, \label{Q-a=0-BOOK}
\end{eqnarray}
which is the expression \eqref{Q-alpha=0-BOOK} for ${a=0}$, exactly the same as the metric function presented in Eq.~(12.19) of \cite{GriffithsPodolsky:2009} for the case ${\epsilon=+1}$ (with ${{\tilde g}=0}$).

It will be shown below that the charged Taub--NUT--(anti-)de Sitter spacetime \eqref{metric-a=0} is \emph{nonsingular} (its curvature does not diverge at ${r=0}$), away from the axis ${\theta=\pi}$ (where the rotating cosmic string is located) it is asymptotically (anti-)de Sitter, and the interior of the black hole is located between its two horizons, that can be surrounded by two ``outer'' cosmological horizons.

\subsection{Uncharged accelerating Kerr--NUT--(anti-)de Sitter black holes\\
 (${e=0=g}$\,: vacuum with $\Lambda$)}

Another nice feature of our new metric \eqref{newmetricGP2005}--\eqref{newQ} is that it \emph{has the same form for vacuum spacetimes} without the electromagnetic field. Indeed, the electric and magnetic charges $e$ and $g$, which generate the electromagnetic field, \emph{enter only the expressions for}~$r_{\Lambda\pm}$ introduced in \eqref{r+}. In other words, $e$ and $g$ just change the values of these two constant parameters. In such vacuum case, they simplify to
\begin{equation}
r_{\Lambda \pm} \equiv \mu \pm \sqrt{\mu^2 + l^2 - a^2 - \lambda}\,. \label{r+-vacuum}
\end{equation}
The metric \eqref{newmetricGP2005}--\eqref{newQ} with \eqref{r+-vacuum} represents the full class of accelerating Kerr--NUT--(anti-)de Sitter black holes. It reduces to accelerating Kerr--(anti-)de Sitter black hole when ${l=0}$, and non-accelerating Kerr--NUT--(anti-)de Sitter black hole when ${\alpha=0}$. For ${a=0}$ it simplifies \emph{directly} to the Taub--NUT--(anti-)de Sitter black hole \eqref{metric-a=0} without acceleration and charges.


\section{Physical analysis of the new metric}
\label{sec_discussion}

The explicit new metric form \eqref{newmetricGP2005}--\eqref{newQ} (or, more generally, \eqref{finalPagain}--\eqref{finalQagain}) of the complete class of accelerating Kerr--Newman--NUT--(anti-)de Sitter black holes is very convenient for investigation of geometric and physical properties of this large family of black holes. This will now be demonstrated by deriving and presenting some of the key quantities and facts, namely those concerning the global structure of the spacetime, the stringy sources of the acceleration, and thermodynamic properties.

\subsection{Curvature of the gravitational field and the electromagnetic field}
\label{subsec:curvature}

First, it is necessary to determine the \emph{gravitational field}, namely the specific curvature of the geometry. It is encoded in the corresponding Newman--Penrose (NP) scalars, that is, components of the curvature tensors with respect to the null tetrad. Its most natural choice~is
\begin{eqnarray}
\textbf{k} \rovno \frac{1}{\sqrt{2}}\, \frac{\Omega}{\rho} \bigg[ \frac{1}{\sqrt{Q}}
 \Big(\big(r^2+(a+l)^2\big)\, \partial_t + a \, \partial_\varphi \Big) + \sqrt{Q} \, \partial_r \bigg] \,, \nonumber \\
\textbf{l} \rovno \frac{1}{\sqrt{2}}\, \frac{\Omega}{\rho} \bigg[ \frac{1}{\sqrt{Q}}
 \Big(\big(r^2+(a+l)^2\big)\, \partial_t + a \, \partial_\varphi \Big) - \sqrt{Q} \, \partial_r \bigg] \,,  \label{nullframe}\\
\textbf{m} \rovno \frac{1}{\sqrt{2}}\, \frac{\Omega}{\rho} \bigg[
 \frac{1}{\sqrt{P} \sin \theta} \Big( \partial_\varphi + \big(a\sin^2\theta +4l\sin^2\!\tfrac{1}{2}\theta \big)\, \partial_t \Big)
 + \mathrm{i} \, \sqrt{P} \, \partial_\theta \bigg] \,. \nonumber
\end{eqnarray}
A direct calculation shows that the only nontrivial Newman--Penrose scalars corresponding to the Weyl tensor and the Ricci tensor are
\begin{eqnarray}
\Psi_2 \rovno \frac{\Omega^3}{\big[r+\mathrm{i}\,(l+a \cos \theta) \big]^3} \bigg[ -(m+\mathrm{i}\,l)\Big(1-\mathrm{i}\,\alpha\, a\,\frac{a^2-l^2}{a^2+l^2}\Big)
- \mathrm{i}\, \frac{\Lambda}{3} \,l \,(a^2-l^2) \nonumber\\
&& \hspace{25mm}
+\frac{(e^2+g^2)}{r-\mathrm{i}\,(l+a \cos \theta)}
\Big(1+\frac{\alpha\,a}{a^2 + l^2} \big[a\,r \cos \theta +\mathrm{i}\, l\, (l+a \cos \theta) \big]\Big)
 \bigg]\,,
 \label{Psi2}\\
\Phi_{11} \rovno \tfrac{1}{2}(e^2+g^2) \,\frac{\Omega^4}{\rho^4} \,,
\label{Phi11}
\end{eqnarray}
respectively, where
\begin{equation}
\Omega  = 1-\frac{\alpha\,a}{a^2+l^2}\, r\,(l+a \cos \theta) \,, \qquad\qquad
\rho^2  = r^2+(l+a \cos \theta)^2 \,, \label{newrho-rep}\\
\end{equation}
cf. \eqref{newOmega}, \eqref{newrho}. The Ricci scalar is simply
\begin{eqnarray}
R \rovno 4\,\Lambda\,,  \label{Ricci-scalar}
\end{eqnarray}
which is the usual relation valid for any solution of Einstein--Maxwell equations with a cosmological constant $\Lambda$. While $\Phi_{11}$ is independent of $\Lambda$, the Weyl curvature component $\Psi_2$ contains the term proportional to ${\Lambda\, l \,(a^2-l^2)}$. The dependence of $\Psi_2$ on the cosmological constant thus disappears if (and only if) ${l=0}$ or ${l=\pm a}$.

\vspace{2mm}

For an invariant identification of curvature singularities and regions which asymptotically become conformally flat, it is  necessary to evaluate the key (second-order) \emph{scalar invariants}, namely the \emph{Kretschmann invariant} ${\mathcal{K}}$ and the \emph{Weyl invariant} ${\mathcal{C}}$,
\begin{eqnarray}
\mathcal{K} \equi  R_{abcd}\, R^{abcd} \,,  \label{WeylWey-def}\\
\mathcal{C} \equi  C_{abcd}\, C^{abcd} \,. \label{RienannRiemann-def}
\end{eqnarray}
This can be conveniently achieved in the NP formalism. Indeed, it is well known that
\begin{equation}
 C^*_{abcd} \, C^{*\, abcd} = 32 \,( \Psi_0 \Psi_4 -4 \Psi_1 \Psi_3 +3 \Psi_2^2 ) \,, \label{Weyl-invariant1}
\end{equation}
in which ${C^*_{abcd} \equiv C_{abcd} + \mathrm{i} \, C^{\sim}_{abcd}}$, where $C^{\sim}_{abcd}$ is the dual tensor to Weyl,
${C^{\sim}_{abcd} \equiv \frac{1}{2} \, \epsilon_{cdef} \, C^{\,\,\,\,\, ef}_{ab}}$. Since ${C^{\sim}_{abcd} \, C^{\sim \,abcd} = - C_{abcd} \, C^{abcd}}$, we get ${C_{abcd} \, C^{abcd} + \mathrm{i} \, C^{\sim}_{abcd} \, C^{abcd} =  \tfrac{1}{2}\,C^*_{abcd} \, C^{* abcd}}$,
see e.g. \cite{Stephanietal:2003}, or Eq.~(17) in \cite{Cherubinietal:2002}. Therefore, the  Weyl invariant is
\begin{equation}
{\mathcal{C} = 16\,\, {\cal R}e\,(\Psi_0 \Psi_4 -4 \Psi_1 \Psi_3 +3 \Psi_2^2)}\,. \label{Weyl-invariant2}
\end{equation}

From the definition of the Weyl tensor it follows that the Kretschmann invariant reads
\begin{equation}
\mathcal{K}  = \mathcal{C} + 2 R_{ab} R^{ab} - \tfrac{1}{3} R^2 \,, \label{Riemann-invariant1}
\end{equation}
where ${R = 4\,\Lambda}$, while $R_{ab} R^{ab}$ can be expressed as\footnote{There are 9 independent (real) quantities encoded in the complex NP scalars ${\Phi_{AB}=\bar\Phi_{BA}}$. Due to their usual definition, the projections on the null tetrad \eqref{nullframe} of the Ricci tensor $R_{ab}$ and of the related traceless Ricci tensor ${S_{ab}\equiv R_{ab}-\frac{1}{4}R\,g_{ab}}$ give the same results. The additional 10th independent component of $R_{ab}$ is given by $\frac{1}{4}R\,g_{ab}$ containing the Ricci scalar $R$, so that ${R_{ab} R^{ab}}$ also involves the term ${\frac{1}{16}R^2\,g_{ab}\,g^{ab}=\frac{1}{4}R^2}$.}
\begin{equation}
\tfrac{1}{8} R_{ab} R^{ab}  = \Phi_{00} \Phi_{22} +  \Phi_{02}\bar\Phi_{02} - 2(\Phi_{01} \bar\Phi_{12}+\bar \Phi_{01} \Phi_{12}) + 2 \Phi_{11}^2  + \tfrac{1}{32}R^2\,. \label{Riemann-invariant2}
\end{equation}

For the \emph{black hole spacetimes} \eqref{newmetricGP2005}--\eqref{newQ}, which are of algebraic type~D, the only nontrival NP scalars are  $\Psi_2$ and $\Phi_{11}$, as given by \eqref{Psi2} and \eqref{Phi11}, respectively. Therefore, the corresponding scalar curvature invariants are
\begin{eqnarray}
\mathcal{C} \rovno 48 \,\, {\cal R}e\,(\Psi_2^2) \,,  \label{WeylWeyl}\\
\mathcal{K} \rovno \mathcal{C} + 32\, \Phi_{11}^2  + \tfrac{8}{3} \Lambda^2 \,. \label{RienannRiemann}
\end{eqnarray}
Interestingly the Weyl invariant takes the explicit factorized form
\begin{equation}
\mathcal{C} = 48 \, \frac{\Omega^6}{\rho^{12}}\, C_+\,C_- \,, \label{Kretschmann}
\end{equation}
where
\vspace{-6mm}
\begin{eqnarray}
\hspace{5mm}C_\pm \rovno m\, \bigg( F_\pm \pm \alpha\,a\,\frac{a^2-l^2}{a^2+l^2}\, F_\mp \bigg) \nonumber\\
&&\hspace{-3mm} \mp\, l\, \bigg( \big[\,1+\tfrac{1}{3}\Lambda\,(a^2-l^2)\big]\,F_\mp  \,\mp \, \alpha\,a\,\frac{a^2-l^2+e^2+g^2}{a^2+l^2}\, F_\pm \bigg) \nonumber\\
&&\hspace{-3mm} - (e^2+g^2)\,\bigg(1+\frac{\alpha\,a}{a^2+l^2}\,r L\bigg) \, T_\pm \,,
\end{eqnarray}
in which ${F_\pm = \big(r \mp L \big) \big( r^2 \pm 4 r L + L^2 \big )}$,
${T_\pm = \big( r^2 \pm 2 r L - L^2 \big)}$, and ${L = l+a \cos \theta}$.

This is a generalization of the previously known expressions for the Kerr--Newman geometry, see \cite {Cherubinietal:2002, LakeZannais:2015} and elsewhere, in which case ${\Lambda, l, g, \alpha=0}$ so that ${\Omega=1}$, ${\rho^2  = r^2 + a^2 \cos^2 \theta}$, and
${C_\pm =  m\,\big(r \mp a \cos \theta \big) \big( r^2 \pm 4 a r \cos \theta + a^2 \cos^2 \theta \big ) - e^2\,\big(r^2 \pm 2 a r \cos \theta - a^2 \cos^2 \theta\big)}$.

\vspace{2mm}

The spacetime also contains \emph{electromagnetic field} represented by the Maxwell tensor $F_{ab}$, forming a 2-form ${\bF = \tfrac{1}{2}F_{ab}\, \dd x^a \wedge \dd x^b =\dd\bA}$. Its 1-form potential ${\bA = A_a \dd x^a}$~is
\begin{equation}
 \bA =  - \frac{e\,r + g\,(l+a \cos \theta)}{r^2+(l+a \cos \theta)^2}\, \dd t  
 + \frac{(e\,r + g\,l)\,(a \sin ^2 \theta + 4 l \sin ^2 \tfrac{1}{2}\theta)
+ g\,\big(r^2+(a+l)^2\big)\cos \theta}{r^2+(l+a \cos \theta)^2} \,\dd\varphi\,.
\label{vector-potential}
\end{equation}
Therefore, the non-zero components of ${F_{ab} = A_{b,a}-A_{a,b}}$ are
\begin{eqnarray}
F_{rt}      \rovno \rho^{-4}\,\Big[\, e\,\big( r^2-(l+a \cos \theta)^2 \big) + 2\,g\,r\,(l+a \cos \theta) \Big]\,, \nonumber \\
F_{\varphi \theta} \rovno \rho^{-4}\,\Big[\, g\,\big( r^2-(l+a \cos \theta)^2 \big) - 2\,e\,r\,(l+a \cos \theta) \Big] \big(r^2 + (a+l)^2\big)\sin \theta  \,, \label{Fab}\\
F_{\varphi r} \rovno \big( a \sin ^2 \theta + 4 l \sin ^2 \tfrac{1}{2}\theta \big)\, F_{rt} \,, \nonumber \\
F_{\theta t} \rovno \frac{a}{r^2 + (a+l)^2}\,F_{\varphi \theta} \,. \nonumber
\end{eqnarray}
The corresponding Newman--Penrose scalars are ${\Phi_0 \equiv F_{ab}\, k^a m^b =  0}$,
${\Phi_2 \equiv F_{ab}\, \bar{m}^a l^b =0 }$, and
\begin{equation}
\Phi_1 \equiv \tfrac{1}{2} F_{ab}( k^a l^b + \bar{m}^a m^b)
   = \frac{\frac{1}{2}(e+\mathrm{i}\,g) \,\Omega^2}{\big(r+\mathrm{i}\,(l+a \cos \theta) \big)^2} \,.
   \label{Phi1}
\end{equation}
It follows that ${\Phi_{11} = 2\,\Phi_1 \bar\Phi_1}$, in fully agreement with \eqref{Phi11}.
 The electromagnetic field thus vanishes if (and only if) ${e=0=g}$.

Since the only nontrivial NP Weyl scalar is $\Psi_2$, \emph{both vectors $\textbf{k}$ and $\textbf{l}$ are principal null directions} (PNDs). In fact, both are double-degenerate, demonstrating that the \emph{gravitational field is of algebraic type D}. The electromagnetic field is non-null, and double-aligned with these PNDs because the only nonzero NP Maxwell scalar is $\Phi_{1}$.

Moreover, by evaluating the \emph{spin coefficients} for the null tetrad \eqref{nullframe} one obtains
\begin{eqnarray}
\kappa  \rovno \nu = 0\,, \qquad \sigma = \lambda = 0\,, \nonumber\\
\varrho \rovno \mu = -\frac{\sqrt{Q}}{\sqrt{2}\, \rho^3} \,
\Big( 1+\mathrm{i} \, \frac{\alpha\, a}{a^2 + l^2}\,(l+a \cos \theta)^2 \Big) \big(r-\mathrm{i}\,(l+a \cos \theta) \big) \,, \label{spincoef} \\
\tau \rovno \pi = -\frac{a\, \sqrt{P} \, \sin \theta}{\sqrt{2}\, \rho^3} \,
\Big( 1-\mathrm{i} \, \frac{\alpha\, a}{a^2 + l^2}\, r^2 \Big) \big(r-\mathrm{i}\,(l+a \cos \theta) \big) \,. \nonumber
\end{eqnarray}
Also ${\alpha = \beta}$ and ${\epsilon = \gamma}$ are non-zero, but we do not write them here due to their complexity.
\newpage

Both double-degenerate PNDs generated by $\textbf{k}$ and $\textbf{l}$ \eqref{nullframe}  are thus \emph{geodetic} (${\kappa=0=\nu}$) and \emph{shear-free} (${\sigma=0=\lambda}$). However, they have \emph{expansion}~$\Theta$ and \emph{twist}~$\omega$ defined, respectively, by the real and imaginary parts of
${\varrho \equiv - (\Theta + \mathrm{i}\, \omega) \equiv \mu}$, namely
\begin{eqnarray}
\Theta \rovno \frac{\sqrt{Q} }{\sqrt{2} \,\rho^3} \,\Big(r+\frac{\alpha\, a}{a^2+l^2}\,(l+a \cos \theta)^3\Big)\,,\label{ThetaPD}\\
\omega \rovno - \frac{\Omega \sqrt{Q} }{\sqrt{2} \,\rho^3} \, (l+a \cos \theta)\,.\label{omegaPD}
\end{eqnarray}

It is now immediately seen from \eqref{omegaPD} that:
\vspace{2mm}

\noindent
$\bullet$ {\bf The black-hole spacetime is everywhere non-twisting if (and only if)}
\begin{equation}
 a=0=l \,.  \label{nontwisting-condition}
\end{equation}
In addition, on the horizons identified by ${Q(r)=0}$ (see below) both the expansion and the twist always vanish (${\Theta=0=\omega}$).
\vspace{3mm}

By inspecting the NP scalars \eqref{Psi2}--\eqref{Ricci-scalar} and \eqref{Phi1}, it is also obvious that:
\vspace{2mm}

\noindent
$\bullet$ {\bf The curvature singularities occur if (and only if)}
\begin{equation}
r = 0
\qquad\quad\hbox{and at the same time}\qquad \quad
l+a \cos \theta = 0 \,.  \label{singularity-condition}
\end{equation}
Indeed, \emph{both} these conditions must be satisfied to have ${r+\mathrm{i}\,(l+a \cos \theta)=0}$. With its complex conjugate, this implies
\begin{eqnarray}
\rho^2 \equiv r^2+(l+a \cos \theta)^2 = 0 \,.
\label{Omega=0}
\end{eqnarray}
This agrees with the Weyl scalar \eqref{Kretschmann}.
\vspace{3mm}

\noindent
$\bullet$ {\bf The region of a generic spacetime is conformally flat if (and only if)}
\begin{eqnarray}
\Omega \rovno 0 \,.
\label{Omega=0}
\end{eqnarray}
With this condition, the spacetime is also locally vacuum, c.f. \eqref{Phi11}, with a cosmological constant $\Lambda$. The condition ${\Omega=0}$ thus localizes the asymptotic \emph{(anti-)de Sitter/Minkowski conformal infinity}.
\vspace{2mm}

\noindent
$\bullet$ {\bf In the case when ${m=0=l}$ and also ${e=0=g}$ then ${\Psi_2=0=\Phi_{11}}$, so that}
\begin{eqnarray}
\hbox{the spacetime is everywhere conformally flat and vacuum.}
\label{Omega=0}
\end{eqnarray}
The metric \eqref{newmetricGP2005}--\eqref{def-mu-and-lambda} then represents {\bf de Sitter spacetime} (for ${\Lambda>0}$), {\bf anti-de Sitter spacetime} (for ${\Lambda<0}$), and {\bf Minkowski spacetime} (for ${\Lambda=0}$).

\vspace{3mm}
\noindent
{\bf Curvature of the subclasses of type~D black holes}, summarized in Sec.~\ref{sec_subclasses}, are easily obtained from the general expression \eqref{Psi2} by setting up the corresponding physical parameters to~zero:
\vspace{2mm}

\noindent
$\bullet$ {\bf Kerr--Newman--NUT--(anti-)de Sitter  (${\alpha=0}$\,: no acceleration)}
\begin{eqnarray}
\Psi_2 \rovno \frac{1}{\big[r+\mathrm{i}\,(l+a \cos \theta) \big]^3} \bigg[ -m
-\mathrm{i}\,l\,\big[\,1+ \tfrac{1}{3}\Lambda\, (a^2-l^2) \big]+\frac{e^2+g^2}{r-\mathrm{i}\,(l+a \cos \theta)}
 \,\bigg]\,,
\label{Psi2-alpha=0}
\end{eqnarray}

\noindent
$\bullet$ {\bf Accelerating Kerr--Newman--(anti-)de Sitter  (${l=0}$\,: no NUT)}
\begin{eqnarray}
\Psi_2 \rovno \frac{(1-\alpha\,r\cos\theta )^3}{(r+\mathrm{i}\,a \cos \theta )^3} \bigg[ -m\,(1-\mathrm{i}\,\alpha\, a)  + (e^2+g^2) \frac{1+\alpha\,r \cos \theta}{r-\mathrm{i}\,a \cos \theta}
 \,\bigg]\,,
\label{Psi2-l=0}
\end{eqnarray}

\noindent
$\bullet$ {\bf Charged Taub--NUT--(anti-)de Sitter  (${a=0}$\,: no rotation)}
\begin{eqnarray}
\Psi_2 \rovno - \frac{m + \mathrm{i}\,l \,(1 - \tfrac{1}{3} \Lambda\,l^2 )}{(r+\mathrm{i}\,l)^3}
  + \frac{e^2+g^2}{(r^2+l^2)(r+\mathrm{i}\,l)^2}\,.
\label{Psi2-a=0}
\end{eqnarray}
Observe that the cosmological constant $\Lambda$ appears in the Weyl curvature scalar $\Psi_2$ only if the NUT parameter $l$ is also present.
\newpage

These expressions further simplify if some of the remaining parameters are zero. In particular, the curvature of \emph{Kerr--Newman--(anti-)de Sitter} black hole is obtained from  \eqref{Psi2-alpha=0} if ${l=0}$. The curvature for \emph{generalized C-metric with $\Lambda$} (accelerating charged black holes without rotation) are obtained from  \eqref{Psi2-l=0} when ${a=0}$. The curvature of \emph{Reissner--Nordstr\"om--(anti-)de Sitter} black hole follows from \eqref{Psi2-a=0} when ${l=0}$. The \emph{uncharged} (vacuum) black holes are obtained for ${e=0=g}$.

\subsection{Horizons}
\label{subsec:horizon}

Next step is the investigation of \emph{horizons of the black hole metric} \eqref{newmetricGP2005}, namely their \emph{number}, possible \emph{degeneration}, and \emph{location}. It is immediately seen that the ``radial'' coordinate~$r$ is spatial in the regions where ${Q(r)>0}$, while it is a temporal coordinate where ${Q(r)<0}$. These regions are separated by horizons $\HH$ located at $r_h$ such that
\begin{eqnarray}
Q(r_h) \rovno 0\,,
\label{Q=0}
\end{eqnarray}
where the key metric function $Q(r)$ is explicitly given by expression \eqref{finalQagain}. In the particular ``under-extreme'' case ${\mu^2 + l^2 > a^2 + e^2 + g^2 + \lambda}$, the alternative form of this function \eqref{newQ} with ${r_{\Lambda+}\ne r_{\Lambda-}}$ can be used.

These observations are in accordance with the behaviour of the \emph{determinant of the metric} \eqref{newmetricGP2005} constrained on a constant~$r$ which, due to the identity ${\rho^2=r^2+(a+l)^2 -a\,(a\sin^2\theta +4l\sin^2\!\tfrac{1}{2}\theta \big)}$, is simply
\begin{eqnarray}
\det \big( g_{\mu \nu} \big|_{r = \text{const.}}\big) = - \, \frac{\rho^2} {\Omega^6} \, Q\, \sin^2 \theta \,.
\end{eqnarray}
Such a 3-surface is thus timelike when ${Q>0}$, while it is spacelike  when ${Q<0}$. On any horizon the determinant vanishes (degenerates) due to \eqref{Q=0}.

Moreover, the determinant of the \emph{complete} metric \eqref{newmetricGP2005} reads ${\,\det g_{\mu \nu} = - \, \Omega^{-8} \rho^4\, \sin^2 \theta}$. This indicates non-regularity only at ${\Omega=0}$ (conformal infinity),  ${\rho=0}$ (curvature singularity),  ${Q=0}$ (horizons), and ${\theta=0}$ or ${\theta=\pi}$ (poles/axes with possible cosmic strings).

Since the function $Q(r)$ does not directly enter the Weyl scalar \eqref{Psi2} or the Ricci scalar \eqref{Phi11} --- and thus the invariants $\mathcal{C}$ and $ \mathcal{K}$ given by \eqref{WeylWeyl} and  \eqref{RienannRiemann} --- there is \emph{no curvature/physical obstacle} located at any of the horizons~$r_h$. Explicit extension of the coordinate system across the horizons $\HH$ will be presented in Sec.~\ref{subsec:global}.

\vspace{2mm}

To analyze the number, possible degeneration, and location of the horizons, it is thus necessary to find all root of the equation \eqref{Q=0}. Because the function \eqref{finalQagain} is a \emph{polynomial of the 4th order}, it admits \emph{up to four real roots}. In the \emph{generic black hole spacetime} \eqref{newmetricGP2005} there is thus \emph{four possible horizons} $\HH$. We can call and denote them as follows:
\vspace{4mm}

$\bullet$    two {\bf black-hole horizons} ${\HH_b^\pm}$ located at ${r_b^\pm}$,
\vspace{2mm}

$\bullet$    two {\bf cosmo-acceleration horizons} ${\HH_c^\pm}$ located at ${r_c^\pm}$.
\vspace{4mm}

While the terminology \emph{black-hole horizon} is common and standard,  we hereby introduce a new name \emph{cosmo-acceleration horizon} which combines the usual names for \emph{cosmological} and for \emph{acceleration} horizons. These are mutually combined in this family of spacetimes due to the presence of \emph{both} the acceleration $\alpha$ \emph{and} the cosmological constant $\Lambda$.
\vspace{2mm}

Let us now analyze these horizons explicitly. The generic key metric function $Q(r)$ is the \emph{quartic polynomial} of $r$, namely
\begin{eqnarray}
Q(r) = q_{4} \, r^4 + q_{3} \, r^3+ q_{2} \, r^2 + q_{1} \, r +q_{0} \,, \label{H:Q(r)general}
\end{eqnarray}
where the coefficients are
\begin{eqnarray}
q_{4} \equi - \, \alpha^2 a^2 \, \frac{a^2-l^2}{(a^2+l^2)^2}-\frac{\Lambda}{3} \,, \nonumber\\
q_{3} \equi 2 \, \alpha \, a \, \bigg[\,
\alpha \, a \, m \, \frac{a^2-l^2}{(a^2+l^2)^2} - \frac{l}{a^2+l^2}
- l \, \frac{a^2-l^2}{a^2+l^2} \, \frac{\Lambda}{3} \bigg] \,, \nonumber\\
q_{2} \equi 1 + 4 \, \alpha \, a \, m \, \frac{l}{a^2+l^2}
- \alpha^2 a^2 \, \frac{a^2-l^2}{(a^2+l^2)^2}\,(a^2 - l^2 + e^2 + g^2 )  - (a^2+3l^2) \, \frac{\Lambda}{3} \,, \label{hat-abcde}\\
q_{1} \equi -2m - 2 \, \alpha \, a \,\frac{l}{a^2+l^2}\,(a^2 - l^2 + e^2 + g^2)  \,,  \nonumber\\[2mm]
q_{0} \equi a^2 - l^2 + e^2 + g^2 \,. \nonumber
\end{eqnarray}

The \emph{quartic equation} ${Q(r)=0}$ can have from zero to maximally four explicit real roots $r_h$ corresponding to the horizons. In particular, we may observe that:

\vspace{4mm}

\noindent
$\bullet$    {\bf Maximally four horizons} is the \emph{general case} which will be discussed in detail in subsequent Sec.~\ref{subsec:gberichorizon}. Some of the roots of  \eqref{Q=0} may coincide, resulting in \emph{degenerate} horizons (doubly, triply, or even quadruply).
\vspace{2mm}

\noindent
$\bullet$    {\bf Maximally three horizons} occur in spacetimes with the physical parameters related in such a way that $q_{4}=0$, that is for
\begin{eqnarray}
\frac{\Lambda}{3} = - \alpha^2 a^2 \, \frac{a^2-l^2}{(a^2+l^2)^2} \,. \label{H:hataEq0}
\end{eqnarray}
For these black hole spacetimes the metric function $Q(r)$ reduces to a \emph{cubic function}. Notice that in the case ${l=0}$, this condition is simply ${\alpha^2=-\Lambda/3}$, i.e., a specific relation between the acceleration of the (rotating and charged) black hole and the \emph{negative} cosmological constant (while the complementary case ${a=0}$ requires ${\Lambda=0}$). Further analysis of this case will be presented in our subsequent paper.
\vspace{2mm}

\noindent
$\bullet$    {\bf Maximally two horizons} occur in spacetimes with such parameters that --- in addition to the condition \eqref{H:hataEq0} --- also the second coefficient in \eqref{H:Q(r)general} vanishes, $q_{3}=0$, that is for ${\alpha\, a = 0 \Rightarrow \Lambda=0}$, or for
\begin{eqnarray}
\alpha\,a\,m = l\,\bigg( \frac{a^2+l^2}{a^2-l^2} -\alpha^2 a^2 \, \frac{a^2-l^2}{a^2+l^2} \bigg) \,. \label{H:hatbEq0}
\end{eqnarray}
Equation \eqref{Q=0} is then a quadratic equation ${q_{2} \, r^2 + q_{1} \, r +q_{0}=0}$, from which both horizons $r_h$ can be easily calculated. If ${q_{1}^2 - 4\, q_{2} q_{0}=0}$, these two horizons coincide (it is double degenerate), and for ${q_{1}^2 - 4\, q_{2} q_{0}<0}$ there is no horizon.
\vspace{2mm}

\noindent
$\bullet$    {\bf Maximally one horizon} occurs when both the constraints \eqref{H:hataEq0} and \eqref{H:hatbEq0} are satisfied, and moreover $q_{2}=0$, that is
\begin{eqnarray}
\alpha^2 a^2 \, (e^2 + g^2 )
= (a^2 + 3 l^2)\,\bigg(\frac{a^2+l^2}{a^2-l^2}\bigg)^2 \,. \label{H:hatcEq0}
\end{eqnarray}
The single horizon is then located at
\begin{eqnarray}
r_h = -\frac{q_{0}}{q_{1}}
  = \frac{1}{4\,\alpha\,a\,l} \,\bigg( a^2+3l^2 + \alpha^2a^2\,\frac{(a^2-l^2)^3}{(a^2+l^2)^2}\bigg)\,. \label{H:hatcEq0rh}
\end{eqnarray}
For ${q_{1}=0}$ there is no horizon.
\vspace{4mm}

These three conditions \eqref{H:hataEq0}, \eqref{H:hatbEq0}, and \eqref{H:hatcEq0} characterize very special black-hole spacetimes in which the physical parameters $\Lambda$, $m$, and ${e^2+g^2}$ have particular values in terms of the Kerr-like rotational parameter $a$, NUT parameter $l$, and acceleration $\alpha$.

It is a usual procedure that \emph{the general quartic equation} \eqref{Q=0}, \eqref{H:Q(r)general} can be solved by first dividing it by a nonzero prefactor $q_{4}$ and then performing the substitution
\begin{equation}
r \equiv x - \frac{q_{3}}{4\,q_{4}}\,, \label{Horizons:substitut}
\end{equation}
leading to the \emph{depressed (reduced) quartic equation} without the cubic term,
\begin{eqnarray}
\frac{1}{q_{4}}\,Q(x)= x^4 + \frac{N}{8\,\mathcal{N}^2}\, x^2 - \frac{R}{8\,\mathcal{N}^3}\, x+ \frac{S}{256\,\mathcal{N}^4} = 0\,,
\label{depressedQ}
\end{eqnarray}
where ${\mathcal{N} \equiv -q_{4}}$, the coefficients are
\begin{eqnarray}
N \equi 8\, q_{4} q_{2} - 3\, q_{3}^2 \,, \label{Horizons:P}\\
R \equi 8\, q_{4}^2 q_{1}  -4\, q_{4} q_{3} q_{2} + q_{3}^3 \,, \label{Horizons:R}\\
S \equi 256\, q_{4}^3 q_{0} -64\, q_{4}^2 q_{3} q_{1} + 16\, q_{4} q_{3}^2 q_{2} -3\, q_{3}^4 \,, \label{Horizons:S}
\end{eqnarray}
and the constants $q_{i}$ are explicitly defined by \eqref{hat-abcde}.

Moreover, the \emph{discriminant} $\Delta$ of the general quartic polynomial \eqref{H:Q(r)general} is
\begin{eqnarray}
\Delta &\!\!\!\equiv&\!\!\!
    256\, q_{4}^3 q_{0}^3
   -192\, q_{4}^2 q_{3} q_{1} q_{0}^2
   -128\, q_{4}^2 q_{2}^2 q_{0}^2
   +144\, q_{4}^2 q_{2} q_{1}^2 q_{0}
   -27 \, q_{4}^2q_{1}^4
     \nonumber \\
&&\!\!\!
   +144\, q_{4} q_{3}^2 q_{2}   q_{0}^2
   -6  \, q_{4} q_{3}^2 q_{1}^2 q_{0}
   -80 \, q_{4} q_{3}   q_{2}^2 q_{1}  q_{0}
   +18 \, q_{4} q_{3}   q_{2}   q_{1}^3
   +16 \, q_{4} q_{2}^4 q_{0}
    \nonumber \\
&& \!\!\!
   -4  \, q_{4} q_{2}^3 q_{1}^2
   -27 \, q_{3}^4 q_{0}^2
   +18 \, q_{3}^3 q_{2} q_{1} q_{0}
   -4  \, q_{3}^3 q_{1}^3
   -4  \, q_{3}^2 q_{2}^3 q_{0}
   +   \, q_{3}^2 q_{2}^2 q_{1}^2\,.
    \label{H:DeltaGeneral}
\end{eqnarray}
This is simply related to the discriminant of the depressed quartic function \eqref{depressedQ} via
\begin{eqnarray}
\Delta \rovno \mathcal{N}^6 \, \Delta_{\text{depressed}}\,, \nonumber
\end{eqnarray}
so that \emph{the signs} of $\Delta$ and $\Delta_{\text{depressed}}$ \emph{are the same}.
\vspace{2mm}

In terms of these key quantities $\Delta, N, S$ and $R$, a \emph{complete} analysis and a full description of the number and the possible multiplicity of roots can now be performed. Following \cite{Rees:1922}, we can summarize that:
\vspace{2mm}

{\baselineskip=17pt
\noindent
{\bf For ${\Delta > 0}$}:

The metric function $Q(r)$ has either \emph{4 distinct real roots}, or none, and that depends on:

\noindent
$\qquad \bullet$ {\bf If ${N < 0}$ and ${N^2 > S}$} then \emph{all 4 roots are real and distinct}.

\noindent
$\qquad \bullet$ {\bf If ${N < 0}$ and ${N^2 < S}$} then there are \emph{2 pairs of complex conjugate non-real roots}.

\noindent
$\qquad \bullet$ {\bf If ${N \geq 0}$} then there are also \emph{2 pairs of complex conjugate non-real roots}.
\vspace{2mm}

\noindent
{\bf For ${\Delta < 0}$}:

The function $Q(r)$ has \emph{2 distinct real roots} and \emph{2 complex conjugate non-real roots}.
\vspace{2mm}

\noindent
{\bf For ${\Delta = 0}$}:

This is the only case when the metric function $Q(r)$ has at least one \emph{multiple root}.

The different cases that can occur are:

\noindent
$\qquad \bullet$ {\bf If ${N < 0}$ together with}:

\noindent
$\qquad \qquad \bullet$ {\bf ${N^2 < S}$}: there is \emph{1 real double root} and \emph{2 complex conjugate roots}.

\noindent
$\qquad \qquad \bullet$ {\bf ${N^2 = S}$}: there are \emph{2 distinct real double roots}.

\noindent
$\qquad \qquad \bullet$ {\bf ${N^2 > S}$ and ${N^2 > - 3 S}$}: there is \emph{1 real double root} and \emph{2 distinct simple real roots}.

\noindent
$\qquad \qquad \bullet$ {\bf ${N^2 = -3 S}$}: there is \emph{1 real triple root} and \emph{1 distinct simple real root}.

\noindent
$\qquad \bullet$ {\bf If ${N > 0}$ together with}:

\noindent
$\qquad \qquad \bullet$ {\bf ${S = 0}$}: there is \emph{1 real double root} and \emph{2 complex conjugate roots}.

\noindent
$\qquad \qquad \bullet$ {\bf ${S > 0}$ and ${R \neq 0}$}: there is also \emph{1 real double root} and \emph{2 complex conjugate roots}.

\noindent
$\qquad \qquad \bullet$ {\bf ${S = N^2}$ and ${R = 0}$}: there are only \emph{2 complex conjugate double roots}.

\noindent
$\qquad \bullet$ {\bf If ${N = 0}$ together with}:

\noindent
$\qquad \qquad \bullet$ {\bf ${S > 0}$}: there is \emph{1 real double root} and \emph{2 complex conjugate roots}.

\noindent
$\qquad \qquad \bullet$ {\bf ${S = 0}$} (implying ${R = 0}$): there is \emph{1 real quadruple root} ${x=0}$, that is ${r_h = -\frac{q_{3}}{4 \, q_{4}}}$.

}

\vspace{4mm}
\noindent
This exhausts all the possibilities.

\subsection{The case with two black-hole and two cosmo-acceleration horizons}
\label{subsec:gberichorizon}

We will now concentrate on physically most interesting case in which there are \emph{four distinct real roots}. This may appear only in the case when ${q_{4}\not=0}$ (otherwise there are maximally three horizons), i.e., when the cosmological constant $\Lambda$ is \emph{not} ``finely tuned'' to acceleration $\alpha$ and the two twist parameters $a$ and $l$, that is for
\begin{eqnarray}
\frac{\Lambda}{3} \ne - \alpha^2 a^2 \, \frac{a^2-l^2}{(a^2+l^2)^2} \,. \label{H:hataNOTEq0}
\end{eqnarray}
In particular, we can observe that for ${\Lambda=0}$ there are no non-accelerating or non-rotating black holes (${\alpha\,a=0}$) with four horizons.

In such generic black-hole spacetimes there are \emph{two black-hole horizons} ${\HH_b^+}$ and ${\HH_b^-}$
and also \emph{two  cosmo-acceleration horizons} ${\HH_c^+}$ and ${\HH_c^-}$.
With the assumption that they are generically distinct, we can rewrite the key metric function $Q(r)$ given by \eqref{H:Q(r)general},
\eqref{hat-abcde} in a factorized form as
\begin{eqnarray}
Q(r) \rovno -\mathcal{N} \,\big(r-r_b^+ \big) \big( r-r_b^- \big)\big(r-r_c^+ \big) \big( r-r_c^- \big)\,, \label{newQrep}
\end{eqnarray}
where ${\mathcal{N} \equiv -q_{4}}$ reads
\begin{eqnarray}
\mathcal{N} \rovno \alpha^2 \, a^2 \, \frac{a^2-l^2}{(a^2+l^2)^2}+\frac{\Lambda}{3}\,, \label{Horizons:N}
\end{eqnarray}
while the 4 roots $r_b^+$, $r_b^-$, $r_c^+$, $r_c^-$ localize the {\bf four distinct horizons}, namely
\begin{eqnarray}
&& \HH_b^+ \kern 0.5em \hbox{at} \kern 0.5em   r_b^+ \quad \hbox{is the \bf outer black-hole horizon}, \label{Hbp:rbp}\\[1mm]
&& \HH_b^- \kern 0.5em \hbox{at} \kern 0.5em   r_b^- \quad \hbox{is the \bf inner black-hole horizon},\\[1mm]
&& \HH_c^+ \kern 0.5em \hbox{at} \kern 0.5em   r_c^+ \quad \hbox{is the \bf outer cosmo-acceleration horizon},\\[1mm]
&& \HH_c^- \kern 0.5em \hbox{at} \kern 0.5em   r_c^- \quad \hbox{is the \bf inner cosmo-acceleration horizon}.\label{Hcm:rcm}
\end{eqnarray}


In view of the classification scheme summarized above, this occurs if (and only if)
\begin{eqnarray}
\Delta > 0 \qquad \text{and}\qquad  N<0 \qquad \text{and} \qquad N^2>S \,. \label{Horizons:Condition4roots}
\end{eqnarray}
Moreover, we can assume a {\bf natural ordering} of these horizons as
\begin{equation}
r_c^- < r_b^- < r_b^+ < r_c^+\,,
\label{Horizons:NaturalOrdering}
\end{equation}
so that the cosmological horizons are located ``outside'' the black hole horizons.
Because ${Q(r)<0}$ for all ${r>r_c^+}$ when ${\mathcal{N}>0}$, such an ordering guarantees that these four horizons separate the corresponding {\bf five regions of the spacetime} in such a way that they are, symbolically expressed,
\begin{eqnarray}
\text{time-dependent} < \text{stationary} < \text{time-dependent} < \text{stationary} < \text{time-dependent} \,. \label{Horizons:ConditionOrdering}
\end{eqnarray}
It means, for example, that in the whole range ${ r \in (r_b^+, r_c^+)}$, the coordinate $r$ is spatial. Therefore, the region between the outer black-hole horizon $\HH_b^+$ and the outer cosmo-acceleration horizon $\HH_c^+$ is stationary.

The natural ordering \eqref{Horizons:NaturalOrdering} implying \eqref{Horizons:ConditionOrdering} is present for a large range of values of the cosmological constant $\Lambda$, including ${\Lambda=0}$. In fact, it is a straightforward generalization of the ordering of two black-hole horizons and two acceleration horizons in the family of type D black holes spacetimes without the cosmological constant, see Eq.~(80) in our previous paper \cite{PodolskyVratny:2021}. The ordering \eqref{Horizons:ConditionOrdering} depends on the constraint ${\mathcal{N}>0}$ which, using \eqref{Horizons:N}, reads
\begin{eqnarray}
\alpha^2 \, a^2 \, \frac{a^2-l^2}{(a^2+l^2)^2}+\frac{\Lambda}{3} > 0 \,. \label{Horizons:N-REPEATED}
\end{eqnarray}
In the ${\Lambda=0}$ case, this condition reduces simply to ${|l| < |a|}$, while in the case ${l=0}$ it is
\begin{eqnarray}
\frac{\Lambda}{3} > - \alpha^2  \,. \label{Horizons:ConditionN}
\end{eqnarray}
Notice also that for ${|l| \ge |a|}$ only (a sufficiently large) ${\Lambda>0}$ is admitted.

An \emph{explicit} evaluation of the 4 distinct roots of the metric function $Q(r)$ in the factorized form \eqref{newQrep} in terms of the 7 physical parameters
${m , a , l , e , g , \alpha , \Lambda}$ is quite cumbersome, leading to rather complicated expressions. Nevertheless, it may be useful to present them here. Using a standard procedure of \textit{Wolfram Mathematica 13} one obtains
\begin{eqnarray}
r_b^\pm \rovno \frac{1}{2}\Big( \sqrt{V}  - H  \pm \sqrt{G-2F/\sqrt{V}} \,\, \Big)\,, \label{rb+-}\\[1mm]
r_c^\pm \rovno \frac{1}{2}\Big( \!\!- \sqrt{V}- H  \pm \sqrt{G+2F/\sqrt{V}} \,\, \Big)\,, \label{rc+-}
\end{eqnarray}
where
\begin{eqnarray}
V \rovno H^2+\frac{1}{3 \, \mathcal{N}}  \Big[ \, 2X - \big(Z+\mathrm{i} \, \sqrt{Y^3-Z^2} \, \big)^{\frac{1}{3}}
 - \big(Z-\mathrm{i} \, \sqrt{Y^3-Z^2}\, \big)^{\frac{1}{3}} \, \Big]  \,, \label{Horizons:V}\\
H \rovno -\frac{K}{\mathcal{N}} \,,\qquad
G = 3H^2+\frac{2X}{\mathcal{N}} - V \,,\qquad
F = H^3+\frac{2 L }{\mathcal{N}}-\frac{K X}{\mathcal{N}^2} \,,
\end{eqnarray}
and
\begin{eqnarray}
K   \rovno \frac{\alpha \, a}{a^2+l^2} \bigg[ \Big(\frac{\alpha \, a}{a^2+l^2} \, m
- \frac{\Lambda}{3}\,l\Big)(a^2-l^2)-l\bigg] \,,\\
L   \rovno m + \frac{\alpha \, a \, l}{a^2+l^2} \, (a^2-l^2+e^2+g^2)\,,\\
X \rovno 1 +4\,\frac{\alpha \,a\,l}{a^2+l^2}\, m - \alpha^2 a^2 \frac{a^2-l^2}{(a^2+l^2)^2} \, (a^2-l^2+e^2+g^2)- (a^2+3l^2) \, \frac{\Lambda}{3}\,, \\
Y \rovno X^2 + 12 \, K  L -12 \, (a^2-l^2+e^2+g^2) \,\mathcal{N} \,, \label{Horizons:Y}  \\[2mm]
Z \rovno X^3 + 18 \, K  L  X - 54\, L ^2 \mathcal{N} + 18\,(a^2-l^2+e^2+g^2) (3 K ^2+2 \, \mathcal{N} X)    \,. \label{Horizons:Z}
\end{eqnarray}
Although these expression are fully explicit, they are not telling much, and so we prefer to postpone their discussion to our subsequent paper.
For example, it is possible to show that the complicated discriminant \eqref{H:DeltaGeneral} can be nicely expressed as
\begin{equation}
\Delta =\tfrac{4}{27}(Y^3-Z^2)\,.
\end{equation}
The condition ${\Delta=0}$ for the existence of multiple roots thus simplifies to ${Y^3=Z^2}$.


\subsection{Ergoregions}
\label{subsec:ergoregions}

For the generic black hole metric \eqref{newmetricGP2005} the condition
\begin{equation}
g_{tt} \equiv \frac{1}{\Omega^2\rho^2}\, ( P \,a^2\sin^2\theta - Q ) = 0
\label{gtt}
\end{equation}
defines the \emph{boundary of the ergoregions}, that are the surface of infinite redshift and also the stationary limit at which observers on fixed $r$ and $\theta$ cannot ``stand still''. It can be seen that for a vanishing Kerr-like rotation parameter~$a$ such a boundary coincides with a horizon determined by ${Q=0}$, but \emph{for any} ${a\not=0}$ there exists a nontrivial ergoregion between the ${g_{tt}=0}$ boundary and the horizon. Moreover, the existence of ergoregions is related only to the Kerr-like rotation parameter $a$, not to the twist NUT parameter~$l$.

There is an ergoregion associated with \emph{any of the four horizons} ${\HH_b^\pm}$ and ${\HH_c^\pm}$. Indeed, the ergoregion boundary \eqref{gtt} is located at
\begin{equation}
 Q(r_e) = a^2\sin^2\theta  \,P(\theta) \,,
\label{gtt=0}
\end{equation}
where the metric functions $P(\theta)$ and $Q(r)$ are given by
\eqref{finalPagain} and \eqref{finalQagain}, or \eqref{newP} and \eqref{newQ}, respectively. For a fixed value of the angular coordinate $\theta$, the right hand side of (\ref{gtt=0}) is a specific constant. Because the function $Q(r)$ is of the fourth order, it follows that there are (at most) \emph{four} boundaries $r_e$ of the ergoregions in the direction of $\theta$.

From \eqref{gtt=0} it is also obvious that \emph{the ergoregion boundary ``touchess'' the corresponding horizon at the poles} because for ${\theta=0}$ and ${\theta=\pi}$ the condition \eqref{gtt=0} reduces to ${Q(r_e) = 0}$.

\begin{figure}[t]
\centerline{\includegraphics[scale=0.6]{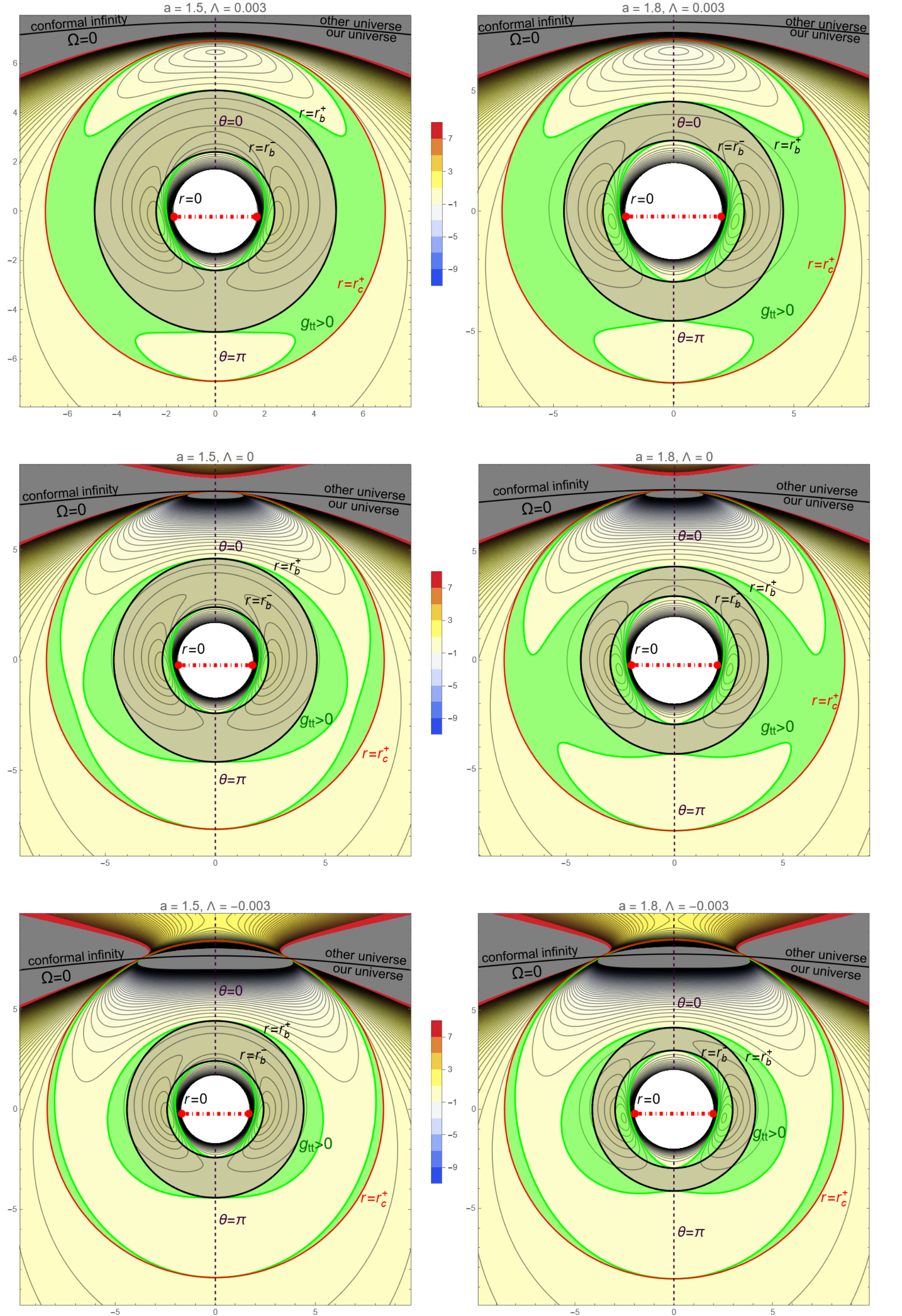}}
\vspace{1mm}
\caption{\small
Plot of the metric function $g_{tt}$ \eqref{gtt} for the generic spacetime \eqref{newmetricGP2005}. The values of $g_{tt}$ are visualized in quasi-polar coordinates ${{\rm x} \equiv \sqrt{r^2 + (a+l)^2}\,\sin \theta}$, ${\mathrm{y} \equiv \sqrt{r^2 + (a+l)^2}\,\cos \theta\,}$ for ${r \geq 0}$. The grey annulus around the center of each figure localizes the black hole bordered by its horizons $\HH_b^\pm$ at ${r_b^+}$ and ${r_b^-}$ (${0<r_b^-<r_b^+}$). The cosmo-acceleration horizon $\HH_c^+$ at ${r_c^+}$ (red circle) and the conformal infinity $\scri$ at $\Omega=0$ are also shown. The grey curves are contour lines ${g_{tt}(r, \theta)=\hbox{const.}}$, and the values are color-coded from red (positive values) to blue (negative values). The green curves are the isolines ${g_{tt}=0}$ determining the boundary of the ergoregions \eqref{gtt=0} in which ${g_{tt}>0}$ (green regions). All six plots are made for the same choice ${m=3}$, ${l=0.2}$, ${e=1.6=g}$, ${\alpha=0.12}$. There are two distinct choices of the Kerr-like rotation parameter, namely ${a=1.5}$ (left) and ${a=1.8}$ (right). The rows visualize three different signs of the cosmological constant, namely
${\Lambda = 0.003}$ (top), ${\Lambda = 0}$ (middle) and ${\Lambda = -0.003}$ (bottom).
For larger values of $a$ and $\Lambda$ the ergoregions are bigger. In fact, the ergoregion above the black hole horizon $\HH_b^+$ \emph{is merged} with the ergoregion below the cosmo-acceleration horizon $\HH_c^+$ in the equatorial part near ${\theta=\frac{\pi}{2}}$.
}
\label{Fig1}
\end{figure}

It is generally complicated to explicitly solve the equation (\ref{gtt=0}), but it can be plotted using a computer. Typical results are shown and discussed in Fig.~\ref{Fig1}.

\vspace{0mm}

\subsection{Curvature singularities}
\label{subsec:singularities}

By inspecting the Newman--Penrose scalars $\Psi_2$ and $\Phi_{11}$ given explicitly as (\ref{Psi2}) and (\ref{Phi11}), we have already concluded that the \emph{curvature singularities occur if and only if} ${\rho^2=0}$, that is when
\begin{equation}
r = 0
\qquad\quad\hbox{and at the same time}\qquad \quad
l+a \cos \theta = 0 \,,  \label{singularity-condition-again}
\end{equation}
see \eqref{singularity-condition}. The presence of these curvature singularities has also been confirmed by the behavior of the Weyl invariant ${\mathcal{C} \equiv  C_{abcd}\, C^{abcd} }$ and the Kretschmann invariant ${\mathcal{K} \equiv  R_{abcd}\, R^{abcd}}$, evaluated in
\eqref{WeylWeyl} and \eqref{RienannRiemann}.

Now, the condition ${l+a \cos \theta = 0}$ \emph{can only be satisfied if} ${|a| \geq |l|}$. Otherwise, ${l+a \cos \theta}$ remains nonzero because ${\cos\theta}$ is bounded to the range $[-1,1]$. Therefore, the curvature singularity structure of the complete family of type~D spacetimes (\ref{newmetricGP2005}) depends on \emph{relative values of the two twist parameters}, that is the Kerr-like rotation parameter $a$ and the NUT parameter $l$, as follows:
\begin{eqnarray}
l=0\,,\ a=0\,:   &\hbox{singularity at}\ r=0 & \hbox{for  any}\ \theta \,, \nonumber\\
l=0\,,\ a\ne0\,: &\hbox{singularity at}\ r=0 & \hbox{for}\ \theta=\pi/2 \,, \nonumber\\
l\ne0\,,\ a=0\,: &\hbox{no singularity\,,}   & \nonumber\\
|l|>|a|>0:       &\hbox{no singularity\,,}   & \label{a-versus-l}\\
l=+a\,:          &\hbox{singularity at}\ r=0 & \hbox{for}\ \theta=\pi\,, \nonumber\\
l=-a\,:          &\hbox{singularity at}\ r=0 & \hbox{for}\ \theta=0\,, \nonumber\\
|a|>|l|>0\,:     &\hbox{singularity at}\ r=0 & \hbox{for}\ \cos \theta = -l/a\,. \nonumber
\end{eqnarray}
These results agree with the well-known character of the ${r=0}$ singularity  of the Schwarzschild--(anti-)de Sitter, Reissner--Nordstr\"{o}m--(anti-)de Sitter and (possibly charged) $C$-metric spacetimes (${l=0}$, ${a=0}$, in this order), the ring singularity structure of the  Kerr--Newman--(anti-)de Sitter black holes (${l=0}$, ${\alpha=0}$), and the absence of curvature singularities in the Taub--NUT--(anti-)de Sitter spacetime (${a=0}$, ${\alpha=0}$).
For a recent detailed analysis of the singular ring structure in these Kerr-like metrics see \cite{ChruscielMaliborskiYunes:2020}.
\newpage

Moreover, from the generic form (\ref{finalQagain}) of the metric function $Q(r)$, or equivalently (\ref{H:Q(r)general}), evaluated at ${r=0}$ we obtain
\begin{eqnarray}
Q(0) \rovno q_{0} \equiv a^2-l^2+e^2+g^2 \,. \label{Q(0)}
\end{eqnarray}
The singularity at ${r=0}$ occurs only if ${a^2 \ge l^2}$, see (\ref{a-versus-l}), so that it is located only in the \emph{stationary} region where ${Q>0}$. In fact, in view of the natural ordering \eqref{Horizons:NaturalOrdering} and the  scheme \eqref{Horizons:ConditionOrdering}, the ring singularity must be contained in the region ${r\in (r_c^- , r_b^-)}$ \emph{between the horizons} $\HH_c^-$ \emph{and} $\HH_b^-$. The alternative possibility ${ r \in (r_b^+ , r_c^+)}$ would correspond to a \emph{naked singularity} in the stationary region located \emph{outside} the horizon $\HH_b^+$.

\subsection{Global structure and conformal diagrams}
\label{subsec:global}

Now we analyze the global structure and the maximal extension of the spacetime. As in the previous parts, we will assume the generic case with \emph{four distinct horizons} $\HH_b^\pm$ and $\HH_c^\pm$ located at ${r_b^\pm}$ and ${r_c^\pm}$, that are ordered as
${r_c^- < r_b^- < r_b^+ < r_c^+}$, see \eqref{Horizons:NaturalOrdering}.

The procedure is basically the same as in Sec.V.D of our previous paper \cite{PodolskyVratny:2021}, and extends special cases of non-accelerating black holes, see e.g. \cite{BoyerLindquist:1967, Carter:1968, HawkingEllis:1973, GibbonsHawking:1977, AkcayMatznera:2011, ChruscielOlzSzybka:2012, LakeZannais:2015, Borthwick:2018}, or black holes with acceleration \cite{GriffithsPodolsky:2006a, GriffithsKrtousPodolsky:2006}.
First, the \emph{retarded} and \emph{advanced null} coordinates are defined,
\begin{eqnarray}
u = t - r_*   \qquad\hbox{and}\qquad   v = t + r_* \,,  \label{uv}
\end{eqnarray}
where the \emph{tortoise} coordinate is
\begin{eqnarray}
r_{*} = \int \frac{r^2+(a+l)^2}{Q(r)} \, \dd r \,, \label{tortoise_r}
\end{eqnarray}
and also the corresponding \emph{untwisted angular} coordinates are introduced by
\begin{eqnarray}
\phi_u = \varphi - a\int\!\frac{\mathrm{d}r}{Q(r)}  \qquad\hbox{and}\qquad
\phi_v = \varphi + a\int\!\frac{\mathrm{d}r}{Q(r)} \,. \label{phiuv}
\end{eqnarray}
Using the \emph{advanced} pair of coordinates $\{v, \phi_v\}$, the metric (\ref{newmetricGP2005}) takes the form
\begin{eqnarray}
\dd s^2 \rovno \frac{1}{\Omega^2}\bigg[\,
\frac{ a^2 P \sin^2 \theta - Q}{\rho^2}\,(\dd v-\T\, \dd \phi_v )^2
+2\,(\dd v-\T\, \dd \phi_v )(\dd r-a\,P \sin^2  \theta \, \dd \phi_v) \nonumber\\
&& \quad +\rho^2 \Big( \frac{\dd \theta^2}{P}+P \sin^2 \theta \, \dd \phi_v^2\Big)\bigg] \,,
\label{metric-advanced}
\end{eqnarray}
where ${\T(\theta) \equiv  a\sin^2\theta +4l\sin^2\!\tfrac{1}{2}\theta}$, while using the \emph{retarded} pair of coordinates $\{u, \phi_u\}$ it reads
\begin{eqnarray}
\dd s^2 \rovno \frac{1}{\Omega^2}\bigg[\,
\frac{a^2 P \sin^2 \theta - Q}{\rho^2}\,(\dd u-\T\, \dd \phi_u )^2
-2\,(\dd u-\T\, \dd \phi_u )(\dd r+ a\,P \sin^2  \theta \, \dd \phi_u) \nonumber\\
&& \quad +\rho^2 \Big( \frac{\dd \theta^2}{P}+P \sin^2 \theta \, \dd \phi_u^2\Big)\bigg] \,.
\label{metric-retarded}
\end{eqnarray}
Both these metrics are \emph{regular} at ${Q(r)=0}$, so that \emph{the coordinate singularities at the horizons has been removed}.

The next step in construction of the maximal (analytic) extension of the manifold is to introduce \emph{both the null coordinates~$u$ and~$v$ simultaneously}, revealing thus the causal structure. The coordinate $r$ is eliminated using the relation \eqref{uv} which implies
\begin{eqnarray}
 2\,\dd r =  \frac{Q}{r^2+(a+l)^2} \,(\dd v-\dd u) \,. \label{dr*uv}
\end{eqnarray}
In addition, it is necessary to construct a \emph{unique angular coordinate} $\phi_h$ across the horizon ar $r_h$ using the specific relation
\begin{eqnarray}
\phi_h = \varphi - \Omega_h \,t\,,
\qquad
\hbox{where}
\qquad
\Omega_h = \frac{a}{r_h ^2 +(a+l)^2} \,.\label{def-phi_h-expl}
\end{eqnarray}
The constant $\Omega_h$ is the \emph{angular velocity of the  horizon}. Actually,
${2\,\dd \phi_h = \dd \phi_u + \dd \phi_v - \Omega_h (\dd u + \dd v)}$.
This it the unique way how to properly combine the distinct angular coordinates $\phi_v$ and $\phi_u$ (for more details see \cite{PodolskyVratny:2021}).

Unfortunately, the specific choice of the angular coordinate $\phi_h$ \emph{depends on the given horizon} via its value $r_h$ and thus $\Omega_h$. For this reason,  it is not possible to find a single and simple global coordinate $\phi$ which would conveniently ``cover'' \emph{all the four horizons}. This drawback was met many years ago already in the Kerr spacetime, so it is not surprising that it reappears in the current context of the complete family of type~D black holes with seven physical parameters.

An explicit \emph{general} metric form of this family constructed in this way reads
\begin{eqnarray}
\dd s^2 \rovno \frac{1}{4\Omega^2}\bigg[
-\frac{Q}{\rho^2}\,\Big((1-\T\Omega_h)(\dd u+\dd v) - 2\T \dd\phi_h\Big)^2
+ Q\rho^2\frac{(\dd u-\dd v)^2 }{[r^2+(a+l)^2]^2}
+ 4\frac{\rho^2}{P}\, \dd \theta^2  \nonumber\\
&& \qquad
+\frac{ P \sin^2 \theta}{\rho^2}\,\Big(\big(a-[r^2+(a+l)^2]\,\Omega_h \big)(\dd u+\dd v) - 2\,[r^2+(a+l)^2]\, \dd\phi_h\Big)^2   \bigg]\,.
\label{metric-doublenull}
\end{eqnarray}
For \emph{non-twisting} black holes without the Kerr-like rotation (${a=0}$) and the NUT parameter (${l=0}$), the metric functions simplify to ${\Omega=1}$, ${P=1}$, ${\rho^2=r^2}$, ${\T=0}$, ${\Omega_h=0}$, so that
\begin{eqnarray}
\dd s^2 \rovno - \frac{Q}{r^2}\,\dd u\,\dd v + r^2(\dd \theta^2 + \sin^2 \theta\,\dd\phi_h^2)\,,
\label{metric-doublenull-nontwist}
\end{eqnarray}
which is the usual form of the spherically symmetric black holes in the double-null coordinates \cite{GriffithsPodolsky:2009}.

It remains to analyze the \emph{global extension of} \eqref{metric-doublenull}  and to study the degree of smoothness (analyticity) of the four distinct horizons $\HH_b^\pm$ and $\HH_c^\pm$ where ${Q(r_h)=0}$.
Restricting to any 2-dimensional section ${\theta=\hbox{const.}}$ and ${\phi_h =\hbox{const.}}$ the general metric \eqref{metric-doublenull}  reduces to
\begin{eqnarray}
\dd \sigma^2 \rovno \frac{1}{4\Omega^2}\bigg[\,
-\frac{(1-\T\Omega_h)^2}{\rho^2}\, Q\,(\dd u+\dd v)^2
+\frac{\rho^2}{[r^2+(a+l)^2]^2}\,Q\,(\dd u-\dd v)^2  \nonumber\\
&& \hspace{8mm}
+\,a^2\,\frac{ P \sin^2 \theta}{\rho^2}\,\frac{(r+r_h)^2(r-r_h)^2}{[r_h^2+(a+l)^2]^2}\,(\dd u+\dd v)^2   \bigg]\,,
\label{metric-doublenull-section}
\end{eqnarray}
which is null at any horizon $r_h$ where ${Q(r_h)=0}$.
Due to the simple \emph{factorized} form \eqref{newQrep} of the metric function $Q(r)$, the integral \eqref{tortoise_r} defining the function $r_*(r)$ can be calculated explicitly as
\begin{eqnarray}
r_{*}(r) =  k_b^+ \, \log \Big| 1-\frac{r}{r_b^+} \Big| + k_b^- \, \log \Big| 1-\frac{r}{r_b^-} \Big|
       + k_c^+ \, \log \Big| 1-\frac{r}{r_c^+} \Big| + k_c^- \, \log \Big| 1-\frac{r}{r_c^-} \Big| \,,
  \label{r*(r)}
\end{eqnarray}
where the auxiliary coefficients are
\begin{eqnarray}
k_b^+ \rovno  - \frac{(r_b^+)^2 + (a+l)^2}{\mathcal{N} \big( r_b^+ - r_b^- \big) \big( r_b^+ - r_c^+ \big) \big( r_b^+ - r_c^- \big) }\,,\nonumber \\
k_b^- \rovno - \frac{(r_b^-)^2 + (a+l)^2}{\mathcal{N} \big( r_b^- - r_b^+ \big) \big( r_b^- - r_c^+ \big) \big( r_b^- - r_c^- \big) }\,,\nonumber \\
k_c^+ \rovno  - \frac{(r_c^+)^2 + (a+l)^2}{\mathcal{N} \big( r_c^+ - r_b^+ \big) \big( r_c^+ - r_b^- \big) \big( r_c^+ - r_c^- \big) }\,,\nonumber \\
k_c^- \rovno  - \frac{(r_c^-)^2 + (a+l)^2}{\mathcal{N} \big( r_c^- - r_b^+ \big) \big( r_c^- - r_b^- \big) \big( r_c^- - r_c^+ \big) }\,.\label{k-hpm}
\end{eqnarray}
Each of these constants is associated with the corresponding horizon $\HH_h^\pm$ located at ${r = r_h^\pm}$, where ${h=b}$ (for the black-hole horizons) or  ${h=c}$ (for the cosmo-acceleration horizons).

We can express the metric functions $Q(r)$, $\rho^2(r)$ and $\Omega^2(r)$ entering \eqref{metric-doublenull-section} in terms of the null coordinates ${v-u}$ instead of~$r$ by using the inversion of the relation ${2\,r_{*}(r)=v-u}$. Finally, we introduce \emph{the couples of new null coordinates} $U_{h}^\pm$ and $V_{h}^\pm$, defined as
\begin{eqnarray}
U_{h}^\pm \rovno (-1)^i\, \sign (k_{h}^\pm) \, \exp \Big(\!\! - \!\frac{u}{2k_{h}^\pm}\Big) \,, \label{U}\\
V_{h}^\pm \rovno (-1)^j\, \sign (k_{h}^\pm) \, \exp \Big(\!\! + \!\frac{v}{2k_{h}^\pm}\Big) \,. \label{V}
\end{eqnarray}
Each couple covers the corresponding horizon $\HH_h^\pm$. Moreover, it is characterized by a \emph{particular choice of two integers} ${(i,j)}$ which specify a certain region in the manifold. Generally, there are \emph{5~types of regions} which are separated by the four types of horizons $\HH_{h}^\pm$, namely

\begin{eqnarray}
\textbf{Region} & \textbf{Description} & \textbf{Specification of $(i,j)$} \nonumber \\
\hbox{I:}   &\hbox{ asymptotic time-dependent domain between }  \HH_{c}^+ \hbox{ and } \scri^+ & (n-2m+1,n+2m-1)\nonumber\\
\hbox{II:}  &\hbox{ stationary region between }  \HH_{b}^+ \hbox{ and } \HH_{c}^+ & (2n-m,2n+m-1) \nonumber\\
\hbox{III:} &\hbox{ time-dependent domain between the black-hole horizons} & (n-2m,n+2m)   \nonumber\\
\hbox{IV:}  &\hbox{ stationary region between }  \HH_{c}^- \hbox{ and } \HH_{b}^-  & (2n-m+1,2n+m) \nonumber\\
\hbox{V:}   &\hbox{ asymptotic time-dependent domain between }  \scri^- \hbox{ and } \HH_{c}^-   & (n-2m+1,n+2m-1) \nonumber
\end{eqnarray}
\vspace{0mm}

\noindent
where $m, n$ are arbitrary integers. The corresponding \emph{Kruskal--Szekeres-type} dimensionless coordinates for every distinct region are
\begin{eqnarray}
T_{h}^\pm = \tfrac{1}{2}(V_{h}^\pm+U_{h}^\pm)\,, \qquad R_{h}^\pm = \tfrac{1}{2}(V_{h}^\pm-U_{h}^\pm) \,.
\end{eqnarray}
(The presence of the \emph{curvature singularity} at ${r=0}$ (implying ${r_*=0}$) for certain values of~$\theta$ restricts the range of the coordinates $U_{b}^-$ and $V_{b}^-$ in the region~IV to the domain outside ${U_{b}^-V_{b}^- = \pm 1}$.)

In terms of these coordinates, the \emph{extension across the horizon is regular} (in fact, analytic). Indeed, by multiplying and dividing the null coordinates \eqref{U} and \eqref{V} we obtain the relations
\begin{eqnarray}
U_{h}^\pm \, V_{h}^\pm \rovno
     \Big( 1-\frac{r}{r_b^+} \Big)^\frac{k_b^+}{k_{h}^\pm}
     \Big( 1-\frac{r}{r_b^-} \Big)^\frac{k_b^-}{k_{h}^\pm}
     \Big( 1-\frac{r}{r_c^+} \Big)^\frac{k_c^+}{k_{h}^\pm}
     \Big( 1-\frac{r}{r_c^-} \Big)^\frac{k_c^-}{k_{h}^\pm}  \,,  \label{UV}\\
\frac{U_{h}^\pm}{V_{h}^\pm} \rovno (-1)^{i+j}\,
     \exp \Big(\!\! - \!\frac{t}{k_{h}^\pm}\Big)  \,,  \label{U/V}
\end{eqnarray}
while the terms ${(\dd u \pm \dd v)^2}$ in the metric \eqref{metric-doublenull-section} become
\begin{eqnarray}
(\dd u \pm \dd v)^2 \rovno \frac{4\,(k_{h}^\pm)^2}{U_{h}^\pm \, V_{h}^\pm}\,
  \bigg(\,\frac{V_{h}^\pm}{U_{h}^\pm}\,(\dd U_{h}^\pm)^2
  \mp 2\, \dd U_{h}^\pm\,\dd V_{h}^\pm
  +\frac{U_{h}^\pm}{V_{h}^\pm}\,(\dd V_{h}^\pm)^2\,\bigg)\,.
\label{du+dv}
\end{eqnarray}
A non-analytic behavior across the horizon $r_h$ may thus occur only at zeros of the product ${U_{h}^\pm \, V_{h}^\pm}$. However, they exactly cancel the zeros of the functions $Q(r)$ in the metric \eqref{metric-doublenull-section}. For example, by choosing the black hole horizon ${r_h=r_b^+}$, we get ${U_{b}^+ \, V_{b}^+ \propto ( r - r_b^+)}$ which obviously compensates the corresponding root ${Q \propto ( r - r_b^+)}$ in \eqref{newQrep}. Notice also that the last term in \eqref{metric-doublenull-section} actually vanishes. Therefore, the metric \eqref{metric-doublenull-section} remains finite at ${r_b^+}$. Of course, the same argument applies to the remaining three horizons.

\emph{Maximal extension} (the complete atlas) of the black-hole manifold represented by \eqref{newmetricGP2005} is obtained by ``glueing together'' the different ``coordinate patches'' \emph{crossing all the horizons}, until a curvature singularity or conformal infinity (the scri $\scri$) is reached. Such an extension has to be performed both along the \emph{advanced} null coordinate~$v$ and the \emph{retarded} null coordinate~$u$, using the corresponding coordinates $U_{h}^\pm$ and $V_{h}^\pm$. By this step-by-step procedure, the coordinate singularities at \emph{all the horizons}  $\HH_h^\pm$ are removed.

Finally, we construct the \emph{Penrose conformal diagrams} visualizing the global structure of this extended manifold. This is achieved by a suitable conformal rescaling of $U_{h}^\pm$ and $V_{h}^\pm$ to the  \emph{compactified} null coordinates $\tilde{u}_{h}^\pm$ and $\tilde{v}_{h}^\pm$ defined as
\begin{eqnarray}
\tan \frac{\tilde{u}_{h}^\pm}{2} \rovno -\sign (k_{h}^\pm) \, (U_{h}^\pm)^{-\sign (k_{h}^\pm)}\,, \\
\tan \frac{\tilde{v}_{h}^\pm}{2} \rovno -\sign (k_{h}^\pm) \, (V_{h}^\pm)^{-\sign (k_{h}^\pm)}\,.
\end{eqnarray}
Consequently, for ${\tilde{T}_{h}^\pm = \frac{1}{2} (\tilde{v}_{h}^\pm+\tilde{u}_{h}^\pm) }$
and
${\tilde{R}_{h}^\pm = \frac{1}{2} (\tilde{v}_{h}^\pm-\tilde{u}_{h}^\pm)}$
we obtain the following explicit expressions in terms of the original coordinates $t,r$ of the metric \eqref{newmetricGP2005}
\begin{equation}
\tilde{T}_{h}^\pm = \left\{
\begin{array}{ll}
      \vspace{0.5em}(-1)^{j+1} \arctan
      \sdfrac{\cosh \frac{t}{2 |k_{h}^\pm |}}{\sinh \frac{r_*}{2 |k_{h}^\pm |}} &
      \text{ for } i+j \text{ even}\,, \\
      \vspace{0.5em}
      (-1)^{j} \arctan \sdfrac{\sinh \frac{t}{2 |k_{h}^\pm |}}{\cosh \frac{r_*}{2 |k_{h}^\pm |}} &
      \text{ for } i+j \text{ odd, } r_* < 0\,, \\
      (-1)^{j} \arctan \sdfrac{\sinh \frac{t}{2 |k_{h}^\pm |}}{\cosh \frac{r_*}{2 |k_{h}^\pm |}} + \pi &
      \text{ for } i+j \text{ odd, } r_* \geq 0\,, \\
\end{array}
\right.\label{TT}
\end{equation}
and
\begin{equation}
\tilde{R}_{h}^\pm =
\left\{
\begin{array}{ll}
      \vspace{0.5em}(-1)^{j} \arctan \sdfrac{\sinh \frac{t}{2 |k_{h}^\pm |}}{\cosh \frac{r_*}{2 |k_{h}^\pm |}} & \text{ for } i+j \text{ even}\,, \\
      \vspace{0.5em}
      (-1)^{j+1} \arctan \sdfrac{\cosh \frac{t}{2 |k_{h}^\pm |}}{\sinh \frac{r_*}{2 |k_{h}^\pm |}} &
      \text{ for } i+j \text{ odd, } r_*<0\,, \\
     (-1)^{j+1} \arctan \sdfrac{\cosh \frac{t}{2 |k_{h}^\pm |}}{\sinh \frac{r_*}{2 |k_{h}^\pm |}} + \pi &
     \text{ for } i+j \text{ odd, } r_* \geq 0\,. \\
\end{array}
\right.\label{RR}
\end{equation}
Recall that the function ${r_*(r)}$ is given by \eqref{r*(r)} and the coefficients $k_{h}^\pm$ by \eqref{k-hpm}. In particular, the lines of constant~$r$ thus coincide with the lines of constant~$r_*$. For every single region the coordinate $r_*$ spans the whole range ${(-\infty,+\infty)}$, and similarly the coordinate~$t$.

\begin{figure}[ht!]
\centerline{\includegraphics[scale=0.5]{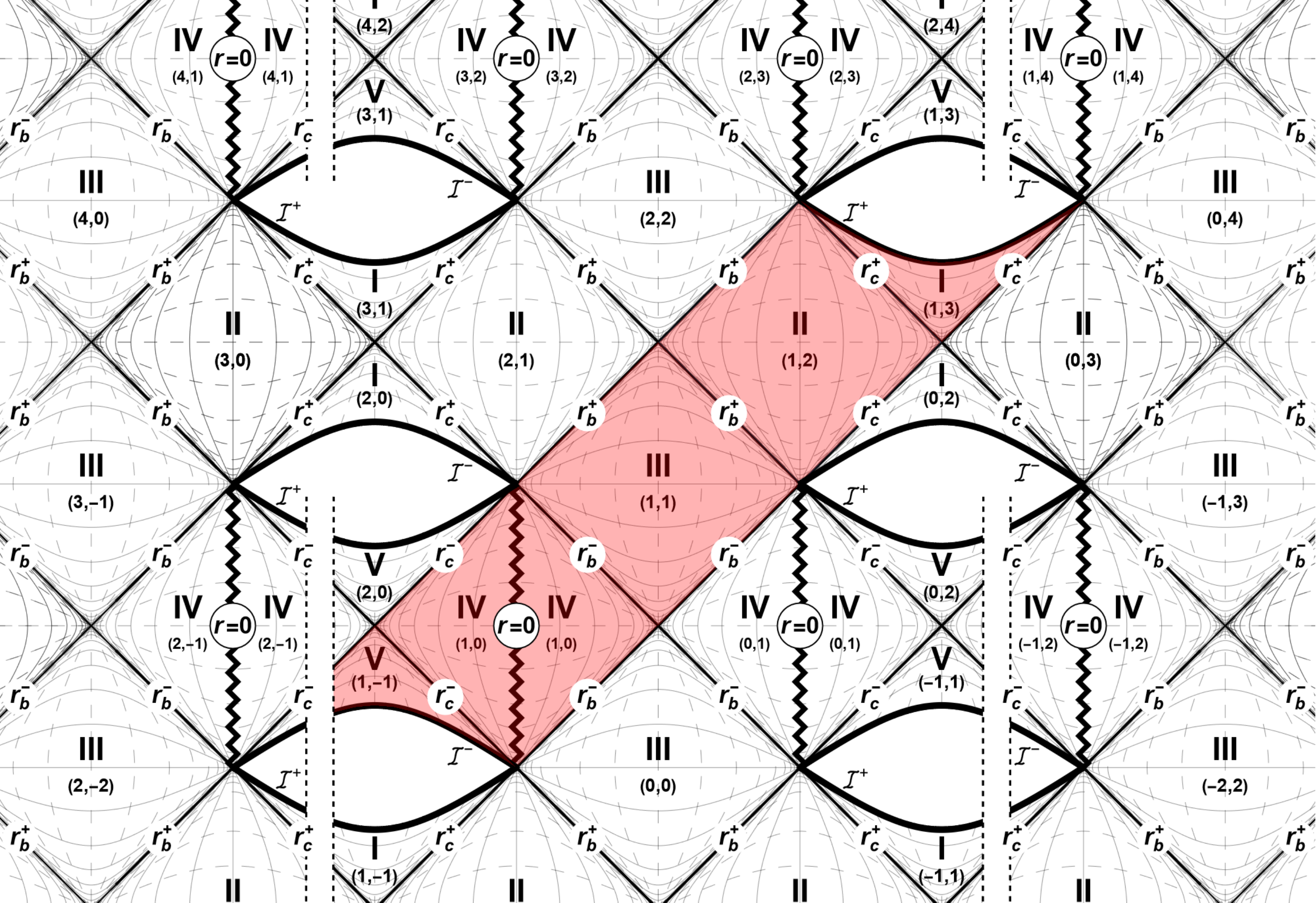}}
\vspace{0mm}
\caption{\small
Penrose conformal diagram of the completely extended spacetime \eqref{newmetricGP2005} showing the global structure of this family of accelerating and rotating charged NUT black holes with a cosmological constant. We assume the ordering of the four distinct horizons as ${r_c^- < r_b^- < r_b^+ < r_c^+}$, see \eqref{Horizons:NaturalOrdering}. Here we show a 2-dimensional section ${\theta, \phi_h =\hbox{const.}}$ with the curvature singularity at ${r = 0}$, i.e., for  ${\theta=\hbox{const.}}$ such that ${\cos \theta = -l/a}$. In such a section, the corresponding regions~IV are ``cut in half'' by this curvature singularity at ${r=0}$, indicated by the vertical zigzag lines. The double dashed vertical parallel lines indicate a separation of distinct asymptotically flat regions close to $\scri^\pm$ (different ``parallel universes'' that are not necessarily identified).
}
\label{Fig2}
\end{figure}

These explicit relations between the compactified coordinates ${\{\tilde{T}_{h}^\pm, \tilde{R}_{h}^\pm \}}$ and the original coordinates ${\{t, r\}}$ of the metric \eqref{newmetricGP2005} for all ${(i,j)}$ can be used for graphical construction of the Penrose diagram, composed of various ``diamond'' regions. The resulting picture is shown in Fig.~\ref{Fig2} for the \emph{special value of}~$\theta$ such that ${\cos \theta = -l/a}$ which \emph{contains the curvature singularity at} ${r = 0}$ in all its regions~IV (see Sec.~\ref{subsec:singularities}). In particluar, for vanishing NUT parameter ${l=0}$ this is the equatorial plane ${\theta=\frac{\pi}{2}}$.

\vspace{0mm}

The complete manifold consists of an \emph{infinite number of the regions}~I, II, III, IV and V, each identified by the specific pair of integers $(i,j)$. These regions are \emph{separated by the corresponding horizons}. Namely, the regions I and II are separated by the cosmo-acceleration horizon ${\HH_c^+}$ at $r_c^+$, with the asymptotic region I also bounded by the conformal infinity $\scri$ (the scri) for very large values of $r$. The regions II and III are separated by the black-hole horizon ${\HH_b^+}$ at ${r_b^+}$, while the regions III and IV are separated by the inner black-hole horizon ${\HH_b^-}$ at ${r_b^-}$. Finally, the regions IV and V are separated by the cosmo-acceleration horizon ${\HH_c^-}$ at $r_c^-$, with the asymptotic region V bounded by the conformal infinity $\scri$ with negative values of $r$. The curves in each region represent the lines of constant~$t$ and~$r$ (dashed or solid, respectively).

In the ``diagonal'' null directions of these Penrose diagrams we can identify the particular coordinate patches covered by the ``advanced'' metric form \eqref{metric-advanced}, extending from the bottom left $\scri^-$ to the top right $\scri^+$ (for example the pink regions~{I--V} between ${(1,-1)}$ and ${(1,3)}$), and also the complementary ``retarded'' metric form  \eqref{metric-retarded}, extending from the bottom right $\scri^-$ to the top left $\scri^+$ (these are not colored but also contain the regions~{I--V}, for example between ${(-1,1)}$ and ${(3,1)}$). These patches ``share'' the ``central regions''~III (for example ${(1,1)}$). Each of such central region III is bounded  by the inner and outer black-hole horizons at $r_b^-$ and $r_b^+$, localizing thus the interior of the corresponding black hole. In the whole extended universe, there are thus \emph{infinitely many black holes} --- they are identified by the different regions~III.

Provided ${|l|\le|a|}$, such black hole has the \emph{curvature singularity} at ${r=0}$ in the region IV bounded by the inner black-hole horizon $\HH_b^-$ at $r_b^-$ (and also the inner cosmo-acceleration horizon $\HH_c^-$ at $r_c^-$). In the section given by the special value of~$\theta$ such that ${\cos \theta = -l/a}$ it is \emph{not possible to cross} from the values ${r>0}$ to ${r<0}$. This is indicated by the vertical zigzag lines in the regions~IV. However, as recently pointed out by MacCallum \cite{MacCallum:2022} in his interesting revisit of the maximal extension of the Kerr black hole spacetime, there is a ``missing triangle'' in usual plots (such as in \cite{GriffithsPodolsky:2009}). Although it is not possible to cross the curvature singularity ${r=0}$ on this specific section, due to its \emph{ring} structure there exist curves that decrease from ${r>0}$ to ${r=0}$ \emph{and continue to} ${r<0}$, provided their value of $\theta$ is \emph{different} form ${\cos \theta = -l/a}$. On such a section there is no curvature singularity, so that the coordinate boundary ${r=0}$ is no obstacle for continuation of the curve. The same argument is valid not only for the Kerr black hole but also for the whole family of rotating black hole spacetimes (such that ${|l|\le|a|}$) investigated here. Therefore, in Fig.~\ref{Fig2} we represent the curvature singularity in (any) region IV simply by a vertical zigzag line. The ``missing triangle'' on the left of ${r=0}$ is the extension of the ``present triangle'' on the right, continuing from positive to negative values of the coordinate~$r$, and vice versa, because the curvature singularity can be ``bypassed'' on any section such that ${\cos \theta \not= -l/a}$.

Each of these black holes, identified by the specific region~III, \emph{is associated with four asymptotic regions}, namely the pair of the regions~I with \emph{future conformal infinity} $\scri^+$  and a pair of the regions~V with \emph{past conformal infinity} $\scri^-$. Moreover, each asymptotically conformally flat region bounded by $\scri$ is \emph{``shared'' by two distinct black holes}. For example, the conformal infinities $\scri^+$ of the \emph{``infinite horizontal chain'' of black holes} (regions~III) given by $\ldots$, ${(3,-1)}$, ${(1,1)}$, ${(-1,3)}$, $\ldots$ are located in the ``future universes'' (regions~I) $\ldots$, ${(5,-1)}$, ${(3,1)}$, ${(1,3)}$, ${(-1,5)}$, $\ldots$, while their ``past universes'' (regions~V) are $\ldots$, ${(3,-3)}$, ${(1,-1)}$, ${(-1,1)}$, ${(-3,3)}$, $\ldots$, respectively. However, these ``past universes'' \emph{need not be the same}. Therefore, we inserted the double dashed vertical parallel lines in them to indicate their separation. Of course, it is possible to ``artificially'' identify (some of) them --- both the black-hole regions~III and/or their asymptotic regions~I and~V. An infinite plethora of various topologically complicated manifolds can thus be constructed.

Let us emphasize that the Penrose conformal diagram shown in Fig.~\ref{Fig2}  represents the global structure of a \emph{generic} black hole spacetime of type D  \eqref{newmetricGP2005} with \emph{4 distinct horizons}. It remains to investigate a great number of other special situations for particular choices of the physical parameters with degenerate (multiple) horizons or with a reduced number of horizons, as identified in Sec.~\ref{subsec:horizon} and Sec.~\ref{subsec:gberichorizon}. Other specific situations also occur, for example $|a|=|l|$. In all these cases the Penrose diagram will have different forms.

\subsection{Regularization of the axes of symmetry ${\theta=0}$ and ${\theta=\pi}$}
\label{subsec:regularization}

As shown in previous works~\cite{GriffithsPodolsky:2005, GriffithsPodolsky:2006, PodolskyVratny:2021}, the metric \eqref{newmetricGP2005} is convenient for explicit analysis of the \emph{regularity of the poles/axes} located at ${\theta=0}$ and ${\theta=\pi}$, respectively, which are the boundaries of the range ${\theta\in[0,\pi]}$.\footnote{Usually, ${\theta=0}$ and ${\theta=\pi}$ are considered as two \emph{semi-axes} of the \emph{same axis of rotation} (a single symmetry axis). This is natural in the simplest spacetimes for which the coordinates ${(r, \theta, \varphi)}$ represent spherical(-like) symmetry with ${r>0}$ only. However, in the present context of generic black hole spacetimes with the Kerr parameter $a$ and the NUT parameter $l$, the range of the ``radial coordinate'' is ${r\in(-\infty, +\infty)}$. In such a case, \emph{both} the axes given by ${\theta=0}$ and ${\theta=\pi}$ have this full range of $r$, and thus they are not the same (unless they are ``artificially'' identified, which would lead to nontrivial topologies). Therefore, they form \emph{two distinct infinite axes} connecting two different asymptotically flat regions in the whole spacetime. This fact is explained in more detail in our previous papers, in particular see Fig.~4 of \cite{PodolskyVratny:2020} and Fig.~2 of \cite{PodolskyVratny:2021}.} This is now further improved with the new metric functions \eqref{newOmega}--\eqref{newQ}.

Recall that there are seven physical parameters in the metric \eqref{newmetricGP2005}, namely $m, a, l, e, g, \alpha, \Lambda$, which represent mass, Kerr-like rotation, NUT parameter, electric and magnetic charges, acceleration, and cosmological constant of the black hole, respectively. But it should be emphasized that, in fact, there is also the \emph{eighth free parameter --- the conicity $C$ hidden in the range of the angular coordinate}
\begin{equation}
\varphi\in[0,2\pi C)\,,
 \label{conicity}
\end{equation}
which has not yet been specified. It is directly related to the \emph{deficit (or excess) angles} of the \emph{cosmic strings (or struts)} located along the axes. The tension associated with these topological defects is the \emph{physical source of the acceleration} of the black holes.

First, let us consider a small circle around the \emph{first axis of symmetry} ${\theta=0}$ in the metric (\ref{newmetricGP2005}) given by ${\theta=\hbox{const.}}$, with the range of $\varphi$ given by \eqref{conicity}, assuming fixed $t$ and~$r$. The invariant length of its circumference is ${\int_0^{2\pi C}\!\! \sqrt{g_{\varphi\varphi}}\, \dd\varphi}$, while its radius is ${\int_0^{\theta}\! \sqrt{g_{\theta\theta}}\, \dd\theta}$, so that
\begin{equation}
 f_0 \equiv \lim_{\theta\to0} \frac{\hbox{circumference}}{\hbox{radius}}
 =\lim_{\theta\to0} \frac{2\pi C \sqrt{g_{\varphi\varphi}} }{ \theta\,\sqrt{g_{\theta\theta}}}  \,.
 \label{Accel-con0}
\end{equation}
For the metric \eqref{newmetricGP2005} near the axis ${\theta=0}$ we get
\begin{equation}
 g_{\varphi\varphi} \approx \frac{P}{\Omega^2\rho^2}\,
    \big(r^2+(a+l)^2\big)^2\,\theta^2 \,,\qquad
 g_{\theta\theta} = \frac{\rho^2}{\Omega^2 P}  \,,
 \label{gphiphi-gthetatheta0}
\end{equation}
and thus, using  \eqref{finalPagain},
\begin{eqnarray}
 f_0 \rovno  2\pi C\,P(0) \label{f0}\\
 \rovno 2\pi C\,
 \Big[ 1 -  2\Big(\,\frac{\alpha\,a\,m}{a^2+l^2} - \frac{\Lambda}{3}\,l \Big)(a+l)
   +\Big(\frac{\alpha^2 a^2}{(a^2+l^2)^2} (a^2-l^2 + e^2 + g^2) + \frac{\Lambda}{3} \Big)(a+l)^2 \Big]\,.
 \nonumber
\end{eqnarray}
Therefore, the axis ${\theta=0}$ in the metric (\ref{newmetricGP2005}) \emph{can always be made regular} by the unique choice of~${C=C_0}$ such that
\begin{eqnarray}
C_0 \equiv \Big[ 1 -  2\Big(\,\frac{\alpha\,a\,m}{a^2+l^2} - \frac{\Lambda}{3}\,l \Big)(a+l)
   +\Big(\frac{\alpha^2 a^2}{(a^2+l^2)^2} (a^2-l^2 + e^2 + g^2) + \frac{\Lambda}{3} \Big)(a+l)^2 \Big]^{-1} \,.
 \label{C0}
\end{eqnarray}
Notice that for ${l=-a}$, this is simply ${C_0=1}$.

Analogously, we can regularize the \emph{second axis of symmetry} ${\theta=\pi}$. By applying the transformation of the time coordinate
\begin{equation}
t_{\pi} \equiv t - 4l\,\varphi\,,
 \label{t-tpi}
\end{equation}
the metric (\ref{newmetricGP2005}) becomes
\begin{align}
\dd s^2 = \frac{1}{\Omega^2} &
  \left(-\frac{Q}{\rho^2}\left[\,\dd t_{\pi}- \left(a\sin^2\theta - 4l\cos^2\!\tfrac{1}{2}\theta \right)\dd\varphi \right]^2
   + \frac{\rho^2}{Q}\,\dd r^2 \right. \nonumber\\
& \quad \left. + \,\frac{\rho^2}{P}\,\dd\theta^2
  + \frac{P}{\rho^2}\,\sin^2\theta\, \big[ a\,\dd t_{\pi} -\big(r^2+(a-l)^2 \big)\,\dd\varphi \big]^2
 \right), \label{newmetricGP2005-Regular-pi}
\end{align}
Now, for ${\theta \to \pi}$ the radius of a small circle around the axis ${\theta=\pi}$  is ${\int_{\theta}^{\pi}\! \sqrt{g_{\theta\theta}}\,\dd\theta}$, so that
\begin{equation}
 f_\pi \equiv \lim_{\theta\to \pi} \frac{\hbox{circumference}}{\hbox{radius}}
 =\lim_{\theta\to\pi} \frac{2\pi C \sqrt{g_{\varphi\varphi}} }{ (\pi-\theta)\,\sqrt{g_{\theta\theta}}}  \,,
 \label{Accel-conpi}
\end{equation}
where for the metric \eqref{newmetricGP2005-Regular-pi} now
 \begin{equation}
 g_{\varphi\varphi} \approx \frac{P}{\Omega^2\rho^2}\,
    \big(r^2+(a-l)^2\big)^2\,(\pi-\theta)^2 \,,\qquad
 g_{\theta\theta} = \frac{\rho^2}{\Omega^2 P}  \,.
 \label{gphiphi-gthetatheta-pi}
\end{equation}
Using \eqref{finalPagain} we obtain
\begin{eqnarray}
 f_\pi \rovno  2\pi C\,P(\pi) \label{fpi}\\
 \rovno 2\pi C\,
 \Big[ 1 +  2\Big(\,\frac{\alpha\,a\,m}{a^2+l^2} - \frac{\Lambda}{3}\,l \Big)(a-l)
   +\Big(\frac{\alpha^2 a^2}{(a^2+l^2)^2} (a^2-l^2 + e^2 + g^2) + \frac{\Lambda}{3} \Big)(a-l)^2 \Big]\,.
 \nonumber
\end{eqnarray}
The axis ${\theta=\pi}$ in the metric (\ref{newmetricGP2005-Regular-pi}) \emph{can always be made regular} by the unique choice ${C=C_\pi}$ where
\begin{eqnarray}
C_\pi \equiv \Big[ 1 +  2\Big(\,\frac{\alpha\,a\,m}{a^2+l^2} - \frac{\Lambda}{3}\,l \Big)(a-l)
   +\Big(\frac{\alpha^2 a^2}{(a^2+l^2)^2} (a^2-l^2 + e^2 + g^2) + \frac{\Lambda}{3} \Big)(a-l)^2 \Big]^{-1} \,.
 \label{Cpi}
\end{eqnarray}
Notice that for ${l=a}$, this is simply ${C_\pi=1}$.

\subsection{Cosmic strings (or struts) and deficit (or excess) angles}
\label{subsec:strings}

Regularizing the second axis ${\theta=\pi}$ by the choice  \eqref{Cpi} there remains a \emph{deficit/excess angle} $\delta_0\equiv 2\pi-f_0$ (conical singularity representing a cosmic string/strut) \emph{along the first axis} ${\theta=0}$, namely
\begin{eqnarray}
 \delta_0 \rovno  \frac{8 \pi a\, \Big[ \alpha a \,  [\,m (a^2+l^2)- \alpha a l
   (a^2-l^2+e^2+g^2)] - \frac{2}{3} \Lambda l (a^2+l^2)^2\Big]}
   {\Big[1+\frac{1}{3}\Lambda(a-l)(a-3 l) \Big](a^2+l^2)^2 +2 \alpha a  m (a-l)
   (a^2+l^2) + \alpha^2a^2 (a-l)^2 (a^2-l^2+e^2+g^2)} \,. \nonumber
 \label{delta0}
\end{eqnarray}

For \emph{nonrotating black holes} (${a=0}$) we immediately obtain ${\delta_0=0}$ which means that \emph{both axes ${\theta=0}$ and ${\theta=\pi}$ are regular}. In such a case, the possible cosmic strings are absent, so that there is \emph{no source of acceleration}. This is fully consistent with our previous observation made in Subsec.~\ref{norotation} that there is  no accelerating ``purely'' NUT--(anti-)de Sitter black hole in the Pleba\'nski--Demia\'nski family of spacetimes. Indeed, by setting the Kerr-like rotation parameter $a$ to zero, the metric \eqref{newmetricGP2005} becomes independent of the acceleration~$\alpha$, and simplifies directly to
\eqref{metric-a=0}.

For black holes \emph{without the NUT parameter} (${l=0}$) this expression simplifies to
\begin{eqnarray}
 \delta_0 \rovno  \frac{8 \pi \alpha  m}{ 1 + 2 \alpha  m + \alpha^2 (a^2+e^2+g^2) +\frac{1}{3}\Lambda a^2 } \,,
 \label{delta0for-l=0}
\end{eqnarray}
recovering the previous results for rotating charged  $C$-metric with a cosmological constant, see Chapter~14 in \cite{GriffithsPodolsky:2009} (and generalizing Eq.~(132) of \cite{PodolskyVratny:2021} to any $\Lambda$). The tension in the cosmic string along ${\theta=0}$ characterized by ${\delta_0>0}$ pulls the black hole, \emph{causing its uniform acceleration}. Such a string extends to the \emph{full range} of the radial coordinate ${r\in (-\infty,+\infty)}$, connecting ``our universe'' with the ``parallel universe'' through the nonsingular black-hole interior close to ${r=0}$.

Complementarily, when the first axis of symmetry ${\theta=0}$ is made regular by the choice (\ref{C0}), there is necessarily an \emph{excess/deficit angle} ${\delta_\pi\equiv 2\pi-f_\pi }$ along the second axis ${\theta=\pi}$, namely
\begin{eqnarray}
 \delta_\pi \rovno  \frac{- 8 \pi a\, \Big[ \alpha a \,  [\,m (a^2+l^2)- \alpha a l
   (a^2-l^2+e^2+g^2)] - \frac{2}{3} \Lambda l (a^2+l^2)^2\Big]}
   {\Big[1+\frac{1}{3}\Lambda(a+l)(a+3 l) \Big](a^2+l^2)^2 -2 \alpha a  m (a+l)
   (a^2+l^2) + \alpha^2a^2 (a+l)^2 (a^2-l^2+e^2+g^2)} \,. \nonumber
 \label{deltapi}
\end{eqnarray}
For ${a=0}$ it gives ${\delta_\pi=0}$, while for ${l=0}$ it simplifies to
\begin{equation}
 \delta_\pi = \frac{ -8\pi \alpha m } { 1 - 2\alpha m +\alpha^2 (a^2+e^2+g^2) +\frac{1}{3}\Lambda a^2  } \,,
 \label{deltapifor-l=0}
\end{equation}
(generalizing Eq.~(134) of \cite{PodolskyVratny:2021} to any $\Lambda$). This represents the \emph{cosmic strut} characterized by ${\delta_\pi<0}$ located along ${\theta=\pi}$ between the pair of black holes, pushing them away from each other in opposite spatial directions.

Interestingly, \emph{both axes} ${\theta=0}$ and ${\theta=\pi}$ can be made \emph{simultaneously regular} (${\delta_0 =0=\delta_\pi}$) if (and only if) seven physical parameters of the black hole spacetime satisfy the special constraint
\begin{eqnarray}
\tfrac{2}{3} \Lambda l (a^2+l^2)^2 = \alpha a \,  \big[\,m (a^2+l^2)- \alpha a l (a^2-l^2+e^2+g^2)\big]  \,.
 \label{bothregularaxes}
\end{eqnarray}
For such a special value of the cosmological constant $\Lambda$, the rotating charged black holes with the NUT parameter ${l\ne0}$ accelerate without the presence of the cosmic strings or struts. In the ${\Lambda=0}$ case the simpler condition given by Eq.~(135) of \cite{PodolskyVratny:2021} is recovered. The condition \eqref{bothregularaxes} also corrects the wrong sign of the $\Lambda$-term in the corresponing unnumbered equation on page 313 of \cite{GriffithsPodolsky:2009}.

\newpage

\subsection{Rotation of the cosmic strings (or struts)}
\label{subsec:rotatingstrings}

With a NUT parameter~${l\ne0}$ these \emph{cosmic strings (or struts) are rotating}. The \emph{angular velocity} parameter~$\omega_\theta$ of the metric (\ref{newmetricGP2005}) is
\begin{equation}
 \omega_\theta \equiv \frac{g_{t\varphi}}{g_{tt}} = -\frac{Q \big(a\sin^2\theta +4l\sin^2\!\tfrac{1}{2}\theta \,\big)
   - a \big(r^2+(a+l)^2\big)P\sin^2\!\theta}
   {Q - a^2 P \sin^2 \theta}\,.
 \label{omega-0}
\end{equation}
Now we consider any fixed value of $r$ away from the horizons (so that ${Q\ne0}$ is a constant). Then the limits ${\theta\to0}$ and ${\theta\to\pi}$ near the two different axes ${\theta=0}$ and ${\theta=\pi}$ give
\begin{equation}
 \omega_0  = 0 \qquad\hbox{and}\qquad  \omega_\pi  = -4l  \,,  \label{omega0pi}
\end{equation}
respectively. The first axis ${\theta=0}$ is thus \emph{non-rotating}, while the second axis ${\theta=\pi}$ rotates, and its \emph{angular velocity is directly (and solely) determined by the NUT parameter~$l$}. Indeed, $\omega_\pi$ does not depend on the Kerr-like parameter~$a$, nor the conicity parameter~$C$. The rotational character of the axis is thus a specific feature related to the NUT parameter~$l$, which is  independent of the possible deficit angles defining the cosmic string/strut along the same axis.

By changing the time coordinate as in \eqref{t-tpi}, we obtain the alternative metric \eqref{newmetricGP2005-Regular-pi} for which
\begin{equation}
 \omega_\theta \equiv \frac{g_{t_{\pi}\varphi}}{g_{t_{\pi}t_{\pi}}}  = -\frac{Q \big(a\sin^2\theta -4l\cos^2\!\tfrac{1}{2}\theta \,\big)
   - a \big(r^2+(a-l)^2\big)P\sin^2\!\theta}
   {Q - a^2 P \sin^2 \theta}\,.
 \label{omega-pi}
\end{equation}
The corresponding angular velocities of the two axes are thus
\begin{equation}
 \omega_0 = 4l \qquad\hbox{and}\qquad  \omega_\pi  = 0  \,.  \label{omega0piregpi}
\end{equation}
In this case, the situation is complementary to \eqref{omega0pi}: the axis ${\theta=0}$ rotates, while the axis ${\theta=\pi}$ does not rotate.

Interestingly, there is a \emph{constant difference}
\begin{equation}
\Delta\omega\equiv\omega_0-\omega_\pi=4l
\label{omega-difference}
\end{equation}
between the angular velocities of the two cosmic strings or struts given by~$l$ (irrespective of the value of $a$ or the choice of $C$). The NUT parameter~$l$ is thus responsible for the \emph{difference} between the magnitude of rotation of the two axes ${\theta=0}$ and ${\theta=\pi}$.

\subsection{Pathological regions with closed timelike curves near the rotating strings (or struts)}
\label{subsec:CTCrotatingstrings}

In the close vicinity of the rotating cosmic strings or struts located along ${\theta=0}$ or ${\theta=\pi}$, the black-hole spacetime  can serve as a time machine because there are closed timelike curves. To identify such ``pathological'' causality-violating regions, let us consider \emph{circles} around the axes of symmetry ${\theta=0}$ or ${\theta=\pi}$ such that only the \emph{periodic} angular coordinate ${\varphi\in[0,2\pi C)}$ changes, while the remaining coordinates $t$,~$r$ and~$\theta$ are constant. The corresponding velocity vectors are thus proportional to the \emph{Killing vector field}~$\partial_\varphi$ whose norm is determined just by the metric coefficient $g_{\varphi\varphi}$ of the general metric \eqref{newmetricGP2005}. There exist regions with
\begin{equation}
g_{\varphi\varphi}<0 \,,
 \label{gphiphi-condition}
\end{equation}
in which the circles (orbits of the axial symmetry) are \emph{closed timelike curves}. Such pathological regions are given by the condition
\begin{equation}
P(\theta) \big(r^2+(a+l)^2\big)^2\sin^2\!\theta   <   Q(r) \big(a\sin^2\theta +4l\sin^2\!\tfrac{1}{2}\theta \,\big)^2 \,,
 \label{gphiphi-general-condition}
\end{equation}
where the functions $P(\theta)$, $Q(r)$ are explicitly given by \eqref{finalPagain}, \eqref{finalQagain}.

Since $P{(\theta)>0}$, this condition can only be satisfied in the regions where ${Q(r)>0}$. In the generic case admitting four distinct horizons \eqref{newQrep}, with ${\mathcal{N}>0}$, ordered as ${r_c^- < r_b^- < r_b^+ < r_c^+}$, the pathological regions with closed timelike curves can only appear in the stationary region ${ r \in (r_b^+, r_c^+)}$ \emph{between the outer black-hole horizon} $\HH_b^+$ \emph{and the outer cosmo-acceleration horizon} $\HH_c^+$, or in the stationary region ${ r \in (r_c^-, r_b^-)}$  \emph{between the inner cosmo-acceleration horizon} $\HH_c^-$ \emph{and the inner black-hole horizon} $\HH_b^-$ containing the curvature singularity at ${r=0}$, see the scheme \eqref{Horizons:ConditionOrdering}. These are, respectively, the regions~II and the regions~IV in the Penrose conformal diagram shown in Fig.~\ref{Fig2}.

Moreover, it can be proven analytically that these \emph{pathological regions with closed timelike curves do not intersect with the ergoregions} (shown in Fig.~\ref{Fig1}), although they are both in the same domains~II and~IV. Indeed, the ergoregions are identified by the condition ${g_{tt}>0}$ (together with ${g_{rr}>0}$), that is
\begin{equation}
 Q < P \,a^2\sin^2\theta \,,
\label{gttergo}
\end{equation}
see Eq.~\eqref{gtt}. Substituting this inequality into  \eqref{gphiphi-general-condition} we obtain
\begin{equation}
 r^2+(a+l)^2 < a^2\sin^2\theta +4al\sin^2\!\tfrac{1}{2}\theta \,.
 \label{pathol-ergo}
\end{equation}
This is the same relation as ${\,r^2+a^2\cos^2\theta+2al\cos\theta+l^2 < 0\,}$, and in view of \eqref{newrho} it reads
\begin{equation}
\rho^2   \equiv r^2+(l+a \cos \theta)^2  <  0  \,,
 \label{pathol-ergo-fin}
\end{equation}
which is a contradiction.

\begin{figure}[ht!]
\centerline{\includegraphics[scale=0.6]{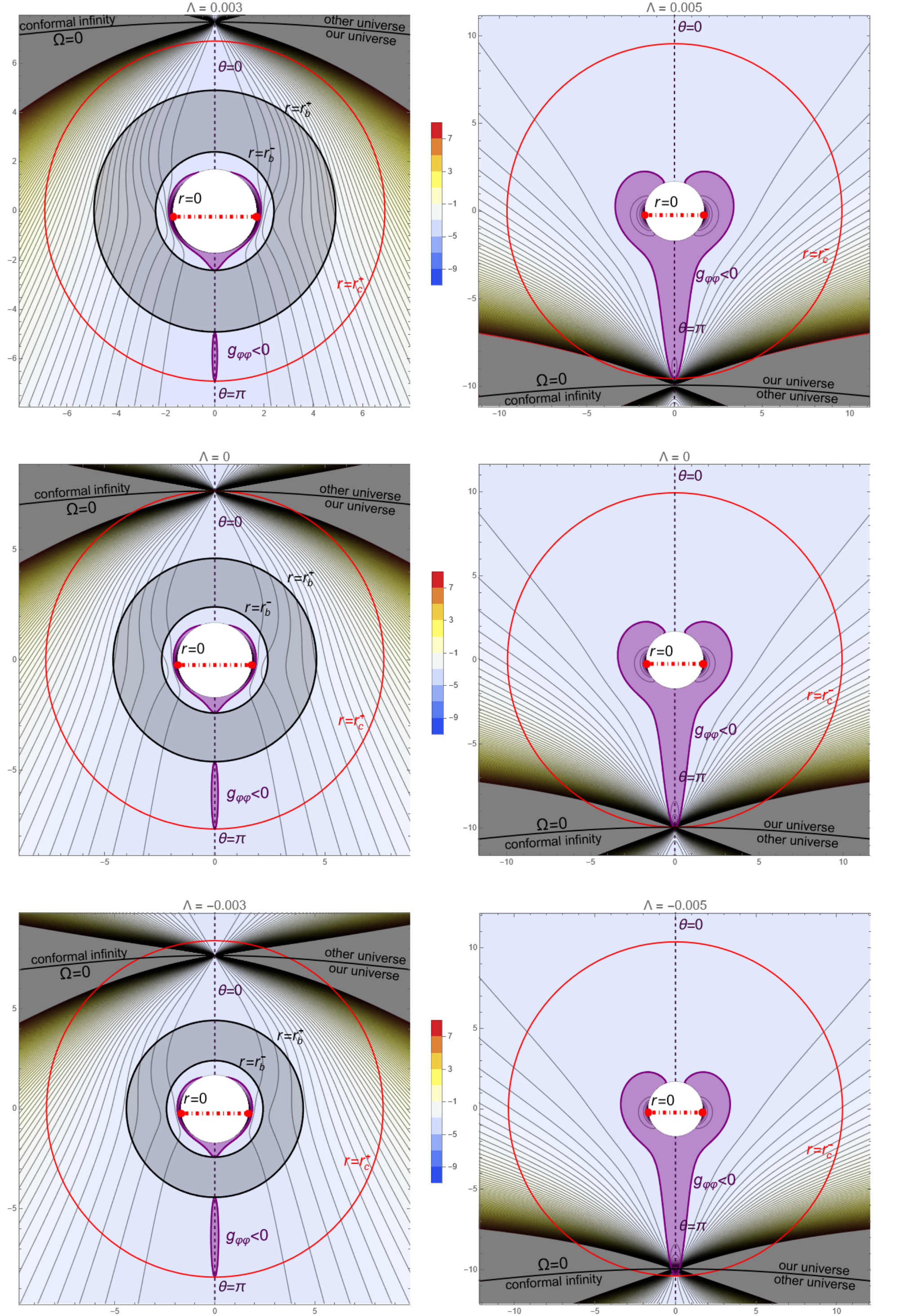}}
\vspace{2mm}
\caption{\small
Plot of the metric function $g_{\varphi\varphi}$ for the accelerating black hole (\ref{newmetricGP2005}) with a regular axis ${\theta = 0}$ and rotating cosmic string (strut) along the axis ${\theta = \pi}$. The values of $g_{\varphi\varphi}$ are visualized in quasi-polar coordinates ${{\rm x} \equiv \sqrt{r^2 + (a+l)^2}\,\sin \theta}$, ${\mathrm{y} \equiv \sqrt{r^2 + (a+l)^2}\,\cos \theta\,}$ for ${r \geq 0}$ (left) and ${r\leq 0}$ (right). The grey annulus in the center of the left figure localizes the black hole bordered by its horizons $\HH_b^+$ at ${r_b^+}$ and $\HH_b^-$ at ${r_b^-}$ (${0<r_b^-<r_b^+}$). The cosmo-acceleration horizons $\HH_c^\pm$ at ${r_c^+}$ and ${r_c^-}$ (big red circles) and the conformal infinity $\scri$ at $\Omega=0$ are also shown. The grey curves are contour lines ${g_{\varphi\varphi}(r, \theta)=\hbox{const.}}$, and the values are color-coded from red (positive values) to blue (negative values); extremely large values are cut. The purple curves are the isolines ${g_{\varphi\varphi}=0}$ determining the boundary of the pathological regions (\ref{gphiphi-general-condition}) with closed timelike curves. They occur close to the axis ${\theta=\pi}$ (purple regions where ${g_{\varphi\varphi}<0}$).  This plot is for the choice ${m=3}$, ${a=1.5}$, ${l=0.2}$, ${e=1.6=g}$, and ${\alpha=0.12}$. The top row is plotted for \emph{positive} values of the cosmological constant (${\Lambda=0.003}$ on the left for ${r \geq 0}$, ${\Lambda=0.005}$ on the right for ${r \leq 0}$), the middle row is for ${\Lambda=0}$, while
the bottom row is plotted for \emph{negative} values of the cosmological constant
(${\Lambda=-0.003}$ on the left for ${r \geq 0}$, ${\Lambda=-0.005}$ on the right for ${r \leq 0}$).
}
\label{Fig3}
\end{figure}

The pathological regions with closed timelike curves are indicated in Fig.~\ref{Fig3} for several choices of the cosmological constant. They are the purple regions near the rotating cosmic string (strut) at  ${\theta=\pi}$.

\vspace{0mm}

\subsection{Thermodynamic quantities}
\label{subsec:thermodynamics}

In this final section we evaluate some basic thermodynamic quantities of the  large class of black holes \eqref{newmetricGP2005}, namely the \emph{entropy}
\begin{equation}
S \equiv \frac{1}{4}\,{\cal A}\,,
\label{entropy}
\end{equation}
given by the horizon area ${\cal A}$, and the \emph{temperature}
\begin{equation}
T\equiv\frac{1}{2\pi}\,\kappa\,,
\label{temperature}
\end{equation}
given by the corresponding horizon surface gravity $\kappa$, see \cite{Wald:book1984}.

The \emph{horizon area} is obtained easily by integrating both angular coordinates of the metric \eqref{newmetricGP2005} for \emph{fixed values of $t$ and ${r=r_h}$},
\begin{equation}
\mathcal{A}(r_h) = \int_0^{2\pi C}\!\!\! \int_{\theta_{\rm min}}^{\theta_{\rm max}} \sqrt{g_{\theta \theta}\, g_{\varphi \varphi}}\,\,\dd \theta \, \dd \varphi\,.
\label{defA}
\end{equation}
Because ${Q(r_h)=0}$ on any horizon, this expression simplifies to
\begin{equation}
\mathcal{A} = 2\pi C\,\big(r_h^2+(a+l)^2\big) \int_{\theta_{\rm min}}^{\theta_{\rm max}}  \frac{\sin\theta}{\Omega^2(r_h)} \,\,\dd \theta \,.
\label{defAsimplified}
\end{equation}
Applying the explicit form of the conformal factor \eqref{newOmega}, that is
\begin{eqnarray}
\Omega(r_h)  \rovno 1-\frac{\alpha\,a\,r_h}{a^2+l^2}\,(l+a \cos \theta) \,, \label{newOmega-again}
\end{eqnarray}
a simple integration leads to
\begin{equation}
\mathcal{A} = 2\pi C\,\big(r_h^2+(a+l)^2\big) \,\frac{a^2+l^2}{\alpha\,a^2 \,r_h}\,
\Big[\, \frac{-1}{\Omega(r_h)} \,\Big]_{\theta_{\rm min}}^{\theta_{\rm max}}  \,.
\label{defAsimplified-withOmega}
\end{equation}

Let us now assume the generic case of \emph{four distinct horizons} $\HH$ introduced in \eqref{Hbp:rbp}--\eqref{Hcm:rcm}. For the {\bf black-hole horizons} $\HH_b^\pm$ the integration range is a full spherical angle, ${[\theta_{\rm min}, \theta_{\rm max}] = [0,\pi]}$, and this leads to the following result:
\begin{eqnarray}
\hbox{area of} \ \HH_b^\pm \ \hbox{is}&
{\displaystyle \mathcal{A}_b^\pm=\frac{ 4\pi C\, \big[(r_b^\pm)^2+(a+l)^2\big]}
{\Big(1-\alpha\, {\displaystyle \frac{a^2+al}{a^2+l^2}}\,r_b^\pm  \Big)\!
 \Big(1+\alpha\, {\displaystyle \frac{a^2-al}{a^2+l^2}}\,r_b^\pm  \Big)} }\,, & \label{Ar+}
\end{eqnarray}
For \emph{vanishing acceleration}~$\alpha$ the area of the black hole horizons is simply
\begin{equation}
\mathcal{A}_b^\pm = 4\pi C\, \big((r_b^\pm)^2+(a+l)^2\big) \,. \label{Ar+-noacceleration}
\end{equation}
This reduces to the well-known expressions for Kerr--Newman--NUT--(anti-)de Sitter  black holes, and in particular the  Schwarzschild solution with a single horizon of the area ${\mathcal{A}_b = 4\pi\, r_{b}^2}$.

Concerning the {\bf cosmo-acceleration horizons} $\HH_c^\pm$, it is necessary to discuss three cases depending on the sign of the cosmological constant. In our previous work \cite{PodolskyVratny:2021} we demonstrated that for ${\Lambda=0}$ the area of both $\HH_a^+\equiv \HH_c^+$ and $\HH_a^- \equiv \HH_c^-$ is \emph{infinite}. The same is true for ${\Lambda<0}$. In this case the reason is that \emph{the cosmo-acceleration horizons extend up to conformal infinity given by} ${\Omega=0}$. This can be seen, e.g., from the corresponding pictures in the bottom row of Fig.~\ref{Fig1} and Fig.~\ref{Fig3} in which $\HH_c^\pm$ are indicated by big red circles. Consequently,
${\Omega(r_c^+,\theta_{\rm min})=0}$ and ${\Omega(r_c^-,\theta_{\rm max})=0}$. In both cases, the expression \eqref{defAsimplified-withOmega} for $\mathcal{A}_c^\pm $ diverges.

For a \emph{positive} cosmological constant ${\Lambda>0}$ the integration \eqref{defAsimplified-withOmega} over the full admitted range ${[\theta_{\rm min}, \theta_{\rm max}] = [0,\pi]}$ implies that
\begin{eqnarray}
\hbox{area of} \ \HH_c^\pm \ \hbox{is}&
{\displaystyle \mathcal{A}_c^\pm=\frac{ 4\pi C\, \big[(r_c^\pm)^2+(a+l)^2\big]}
{\Big(1-\alpha\, {\displaystyle \frac{a^2+al}{a^2+l^2}}\,r_c^\pm  \Big)\!
 \Big(1+\alpha\, {\displaystyle \frac{a^2-al}{a^2+l^2}}\,r_c^\pm  \Big)} }\,, & \label{Arc+}
\end{eqnarray}
Interestingly, \emph{these areas of cosmo-acceleration horizons $\HH_c^\pm$ are finite}.

Indeed, from the general form \eqref{finalQagain} of the metric function $Q(r)$, namely
\begin{eqnarray}
Q(r) \rovno \Big[\,r^2 - 2m\, r  + (a^2-l^2+e^2+g^2) \Big]
            \Big(1+\alpha\,a\,\frac{a-l}{a^2+l^2}\, r\Big)
            \Big(1-\alpha\,a\,\frac{a+l}{a^2+l^2}\, r\Big)\nonumber\\
   &&  - \frac{\Lambda}{3}\,r^2 \Big[\,r^2 + 2\alpha\,a\,l\,\frac{a^2-l^2}{a^2+l^2}\,r + (a^2+3l^2)\,\Big]. \label{finalQagain-final}
\end{eqnarray}
evaluated at the horizons $r_c^\pm$ (which are defined as the two roots of ${Q(r_c)=0}$), it follows that
\begin{eqnarray}
\Big(1-\alpha\,\frac{a^2+al}{a^2+l^2}\, r_c\Big)
\Big(1+\alpha\,\frac{a^2-al}{a^2+l^2}\, r_c\Big)
=  \frac{\Lambda}{3}\,r_c^2\, \frac
{r_c^2 + 2\alpha\,a\,l\,\frac{a^2-l^2}{a^2+l^2}\,r_c + (a^2+3l^2)}
{r_c^2 - 2m\, r_c  + (a^2-l^2+e^2+g^2)}, \label{proof-accel}
\end{eqnarray}
An infinite value of $\mathcal{A}_c^\pm$ given by \eqref{Arc+} would require the left-hand side of \eqref{proof-accel} to be \emph{zero}, implying its roots ${r_c = \pm \frac{1}{\alpha}\,\frac{a^2+l^2}{a^2 \pm a l} }$. By substituting such values into the numerator of the right-hand side of \eqref{proof-accel} we get
${r_c^2 + 2\alpha\,a\,l\,\frac{a^2-l^2}{a^2+l^2}\,r_c + (a^2+3l^2) = \frac{(a^2+l^2)^2}{\alpha^2 a^2 (a\pm l)^2} + (a\pm l)^2 }$ which is \emph{strictly positive}. For ${\Lambda>0}$ we thus get a contradiction, so that $\mathcal{A}_c^\pm$ must be finite.

For ${m=a=l=e=g=\alpha=0}$ (so that ${C=1}$) the function reduces to ${Q(r) = r^2 \,(1 - \frac{\Lambda}{3}\,r^2)}$. The cosmological horizons are thus located at ${r_c^2=\frac{3}{\Lambda}}$, and their areas given by \eqref{Arc+} are $\mathcal{A}_c = 4\pi\, r_c^2 = 12\pi/\Lambda$ which is the well-know result for the de Sitter space.

\vspace{2mm}

The \emph{temperature} of the horizon is determined by its \emph{surface gravity}~$\kappa$. In \cite{MatejovPodolsky:2021, PodolskyVratny:2021} we showed that for the general metric form \eqref{newmetricGP2005} this can be expressed as
\begin{equation}
\kappa = \frac{1}{2}\,\frac{Q'(r_h)}{r_h^2+(a+l)^2}\,,
\label{eq:kappa}
\end{equation}
where the prime denotes the derivative with respect to~$r$. With the factorized form \eqref{newQrep} of the metric function $Q(r)$, using the constant parameters \eqref{k-hpm}, this can be easily evaluated as
\begin{eqnarray}
\hbox{surface gravity of} \ \HH_b^+\, \hbox{is}\quad \kappa_b^+ = \frac{1}{2 k_b^+}\rovno - \frac{\mathcal{N}}{2} \frac{( r_b^+ - r_b^- )( r_b^+ - r_c^+)( r_b^+ - r_c^-)}{(r_b^+)^2 + (a+l)^2} , \label{kappa-b+}\\[1mm]
\hbox{surface gravity of} \ \HH_b^- \ \hbox{is}\quad  \kappa_b^- = \frac{1}{2 k_b^-}\rovno - \frac{\mathcal{N}}{2} \frac{( r_b^- - r_b^+ )( r_b^- - r_c^+)( r_b^- - r_c^-)}{(r_b^-)^2 + (a+l)^2} , \label{kappa-b-}\\[1mm]
\hbox{surface gravity of} \ \HH_c^+ \ \hbox{is}\quad  \kappa_c^+ = \frac{1}{2 k_c^+}\rovno - \frac{\mathcal{N}}{2} \frac{( r_c^+ - r_b^+ )( r_c^+ - r_b^-)( r_c^+ - r_c^-)}{(r_c^+)^2 + (a+l)^2} \,, \label{kappa-a+}\\[2mm]
\hbox{surface gravity of} \ \HH_c^- \ \hbox{is}\quad  \kappa_c^- = \frac{1}{2 k_c^-}\rovno - \frac{\mathcal{N}}{2} \frac{( r_c^- - r_b^+ )( r_c^- - r_b^-)( r_c^- - r_c^+)}{(r_c^-)^2 + (a+l)^2} \,. \label{kappa-a-}
\end{eqnarray}
It can now be seen from \eqref{kappa-b+} and \eqref{kappa-b-} that
\begin{equation}
\kappa_b^+=0=\kappa_b^- \qquad \hbox{if} \qquad r_b^+ = r_b^- \,,
\label{eq:extremal-case1}
\end{equation}
and from \eqref{kappa-b+} and \eqref{kappa-a+} that
\begin{equation}
\kappa_b^+=0=\kappa_c^+ \qquad \hbox{if} \qquad r_b^+ = r_c^+ \,.
\label{eq:extremal-case1}
\end{equation}
This confirms that \emph{extremal horizons have vanishing surface gravity}, and thus zero thermodynamic temperature~${T=\frac{1}{2\pi}\,\kappa}$.

\section{Summary}
\label{sec:summary}

We presented a new metric form \eqref{newmetricGP2005}--\eqref{finalQagain} of the large family of exact black holes of algebraic type~D, initially found by Debever (1971) and by Pleba\'nski and Demia\'nski (1976). It generalizes our previous paper on this topic \cite{PodolskyVratny:2021} to \emph{any value of the cosmological constant}~$\Lambda$. We also demonstrated that this improved metric representation simplify the investigation of various geometrical and physical properties. In particular:

\begin{itemize}

\item In Sec.~\ref{sec_derivation} we recalled the Griffiths--Podolsk\'y (2005, 2006) form of this class of spacetimes, and we further improved it by introducing a modified set of the mass and charge parameters ${m, e, g}$, applying a conformal rescaling $S$, and choosing a gauge of the twist parameter $\omega$.

\item As summarized in Sec.~\ref{sec:introsummary}, the metric \eqref{newmetricGP2005} and its functions \eqref{newOmega}--\eqref{finalQagain} are simple, depending only on the radial coordinate $r$ and the angular coordinate $\theta$. Moreover, the key functions $P(\theta)$ and $Q(r)$ can be further compactified to \eqref{newP}--\eqref{newQ}. In particular, $P(\theta)$ is factorized.

\item The metric depends on seven parameters $m, a, l, e, g, \alpha, \Lambda$ with direct physical meaning. They represent the mass parameter, Kerr-like rotation, NUT parameter, electric and magnetic charges,  acceleration of the black hole, and the cosmological constant, respectively.

\item Another nice feature of the new metric form \eqref{newmetricGP2005}--\eqref{finalQagain} is that any of its seven physical parameters can be independently set to zero (and this can be done in any order). As shown in Sec.~\ref{sec_subclasses}, specific subclasses of type~D black holes are thus easily obtained. These are the black holes with ${\Lambda=0}$, obtained and analyzed previously in \cite{PodolskyVratny:2021}, Kerr--Newman--NUT--(anti-)de Sitter black holes without acceleration (${\alpha=0}$), accelerating Kerr--Newman--(anti-)de Sitter black holes without NUT (${l=0}$), charged Taub--NUT--(anti-)de Sitter black holes without rotation (${a=0}$), and accelerating Kerr--NUT--(anti-)de Sitter black holes without electric or magnetic charges (${e=0}$ or ${g=0}$).

\item All the metric functions \eqref{newOmega}--\eqref{finalQagain} depend on the acceleration $\alpha$ only via the product ${\alpha\,a}$. Consequently, by setting the Kerr-like rotation $a$ to zero, the new metric \eqref{newmetricGP2005} always becomes independent of~$\alpha$, and simplifies directly to the charged Taub--NUT--(anti-)de Sitter black holes. This explicitly confirms the previous observation made by Griffiths and Podolsk\'y that there is no accelerating purely NUT black hole in the Pleba\'nski--Demia\'nski family of type~D spacetimes. Quite surprisingly, such a solution for accelerating non-rotating black hole with just the NUT parameter and ${\Lambda=0}$ exists\cite{ChngMannStelea:2006,PodolskyVratny:2020}, but it is of distinct algebraic type~I. Its possible generalization to any cosmological constant $\Lambda$ remains an open problem.

\item The simplest subcases of the metric \eqref{newmetricGP2005} with just the mass parameter $m$ and a cosmological constant $\Lambda$, plus  one additional physical parameter, give famous black holes, namely the Schwarzschild--(anti-)de Sitter, Reissner--Nordstr\"{o}m--(anti-)de Sitter, Kerr--(anti-)de Sitter, Taub--NUT--(anti-)de Sitter black holes, or black holes accelerating in de Sitter or anti-de Sitter universes --- all in their usual coordinate forms.

\item As shown in Sec.~\ref{sec_discussion}, our convenient metric \eqref{newmetricGP2005}--\eqref{finalQagain} considerably simplifies the study of physical and geometrical properties of this large family of black holes. First of all, the Weyl and Ricci curvature tensors, expressed as the Newman--Penrose scalars $\Psi_2$ and $\Phi_{11}$ (with respect to the natural tetrad \eqref{nullframe} adapted to the double-degenerate principal null directions) can be evaluated, confirming the type~D algebraic structure of the gravitational field, aligned with the non-null electromagnetic field \eqref{vector-potential}--\eqref{Phi1}.

\item  Their form \eqref{Psi2} and \eqref{Phi11}, together with the explicit expressions \eqref{WeylWeyl} and \eqref{RienannRiemann} for the Kretschmann scalar ${\mathcal{K} \equiv R_{abcd}\, R^{abcd}}$ and the Weyl scalar ${\mathcal{C} \equiv C_{abcd}\, C^{abcd}}$, clarifies the presence and the structure of the curvature singularity. It is located at $\rho^2=0$, i.e., at ${r = 0}$, but only if also ${l+a \cos \theta = 0}$, which requires ${|l| \leq |a|}$. There is no curvature singularity in the black-hole spacetimes with large NUT parameter ${|l|>|a|\ge 0}$.

\item Both the double-degenerate principal null directions $\textbf{k}$ and $\textbf{l}$ given by \eqref{nullframe} are geodetic, shear-free, and expanding. They are twisting if and only if ${a=0=l}$.

\item  The generic black-hole spacetime becomes asymptotically conformally flat at the conformal infinity localized by the condition ${\Omega = 0}$.

\item In general, there are four distinct horizons identified by the roots ${Q(r_h)=0}$ of the metric function $Q(r)$ --- which is explicitly given by \eqref{finalQagain} --- a pair of black-hole horizons $\HH_b^\pm$ at ${r_b^\pm}$, and a pair of cosmo-acceleration horizons $\HH_c^\pm$ at ${r_c^\pm}$. The positions of these four horizons are explicitly given by expressions \eqref{rb+-} and \eqref{rc+-}, respectively. Their natural ordering is ${r_c^-<r_b^-<r_b^+<r_c^+}$.

\item Of course, there may be less then four horizons, and they can be degenerate (corresponding to multiple roots of ${Q(r_h)=0}$), as explicitly listed in Subsec.~\ref{subsec:horizon}.

\item Whenever the Kerr-like rotation parameter~$a$ is non-zero, each of these four horizons is accompanied by the corresponding ergoregion, see Subsec.~\ref{subsec:ergoregions} and Fig.~\ref{Fig1}.

\item The ring-like curvature singularity at ${r = 0}$ such that ${\cos \theta = -l/a}$ (requiring ${a^2 \geq l^2}$) is, for the black hole solution, located in the stationary region IV between the inner cosmo-acceleration horizon $\HH_c^-$ and the inner black-hole horizon $\HH_b^-$ (assuming the natural ordering ${r_c^-<r_b^-<r_b^+<r_c^+}$).

\item in Subsec.~\ref{subsec:global} we analyzed the global causal structure of the generic family of black-hole spacetimes \eqref{newmetricGP2005} by constructing the Kruskal--Szekeres-type coordinates which enabled us to perform  the maximal analytic extension across all the horizons. It revealed an infinite number of time-dependent regions (of type I, III, V) and stationary regions (of type II, IV) which are separated by the black-hole and cosmo-acceleration horizons $\HH_b^\pm$ and $\HH_c^\pm$.

\item This global structure is visualized in the Penrose diagrams obtained by a suitable conformal compactification, drawn in Fig.~\ref{Fig2}. The complete manifold contains an infinite number of black holes in various universes identified by distinct (future and past) conformal infinities~$\scri$.

\item In Subsec.~\ref{subsec:regularization} we investigated the regularization of the two axes of axial symmetry ${\theta=0}$ and ${\theta=\pi}$ by an appropriate setting of the conicity parameter $C$ in the range ${\varphi\in[0,2\pi C)}$. The first axis ${\theta=0}$ is regular in the metric form \eqref{newmetricGP2005} with the choice \eqref{C0}, while the second axis ${\theta=\pi}$ is regular in the metric form \eqref{newmetricGP2005-Regular-pi} with the choice \eqref{Cpi}.

\item Both these choices lead to the existence of a cosmic string or a strut identified by the deficit or excess angle on the complementary axis, see the expressions for $\delta_0$ and $\delta_\pi$ in Subsec.~\ref{subsec:strings}. Such topological defects are the physical source of acceleration of the black holes.

\item Interestingly, both the axes of symmetry can be made regular simultaneously for the particular choice  \eqref{bothregularaxes} of the physical parameters.

\item In addition to such deficit/excess angles, the cosmic strings/struts are characterized by their rotation $\omega$ (angular velocity). In Subsec.~\ref{subsec:rotatingstrings} we demonstrated that their values are directly related to the NUT parameter $l$, see the expressions (\ref{omega0pi}) and (\ref{omega0piregpi}). There is always a constant difference ${\Delta\omega=4l}$ between the angular velocities of the two rotating cosmic strings or struts.

\item In the vicinity of these rotating strings/struts there are pathological regions with closed timelike curves, see Subsec.~\ref{subsec:CTCrotatingstrings} and Fig.~\ref{Fig3}.

\item Although the pathological regions with closed timelike curves are located in the same domains as the ergoregions, they do not overlap with each other, see the end of Subsec.~\ref{subsec:CTCrotatingstrings}.

\item The new metric form \eqref{newmetricGP2005} is also convenient for the investigation of thermodynamic quantities. In Subsec.~\ref{subsec:thermodynamics} we evaluated the area and the surface gravity of the black-hole and cosmo-acceleration horizons, simply related to their entropy and temperature.

\end{itemize}

All this demonstrates the usefulness of the new improved metric of the complete family of type~D accelerating and rotating black holes with charges and the NUT parameter in (anti-)de Sitter universe. Various other investigations can now be performed. Among them is a systematic analysis of the degenerate cases with smaller number of horizons, and with multiple horizons. Recently, such extremal isolated horizons have been studied, for example in the works \cite{LewandowskiPawlowski:2003, KunduriLucietti:2009a, KunduriLucietti:2009b, KunduriLucietti:2013, BukLewandowski:2021, MatejovPodolsky:2021, MatejovPodolsky:2022}. Also, extension of the Pleba\'nski--Demia\'nski solutions (including a cosmological constant) to the framework of the metric-affine gravity (MAG) theory was constructed in \cite{BahamondeValcarcelJarv:2022}. It would be nice to see if the new and more explicit metric \eqref{newmetricGP2005}--\eqref{finalQagain} simplifies such investigations.

\section*{Acknowledgments}

This work has been supported by the Czech Science Foundation Grant No.~GA\v{C}R 22-14791S (JP), and by the Charles University project GAUK No.~358921 and the Czech Science Foundation Grant No.~GA\v{C}R 23-05914S  (AV).


\end{document}